\documentclass[useAMS]{mn2e}

\usepackage{graphicx}
\usepackage{amssymb}
\def\max{\mathop{\rm max}\limits}

\title[Hilda collisional family]%
{Did the Hilda collisional family form during the late heavy bombardment?}
\author[M. Bro\v{z} et~al.]{M. Bro\v{z}$^{1}$\thanks{E-mail: mira@sirrah.troja.mff.cuni.cz},
 D. Vokrouhlick\'y$^{1}$,
 A. Morbidelli$^2$,
 D. Nesvorn\'y$^3$,
 W.F. Bottke$^3$
 \\
$^{1}$Institute of Astronomy, Charles University, Prague, V Hole\v sovi\v ck\'ach 2, 18000 Prague 8, Czech Republic\\
$^{2}$Observatoire de la Côte d'Azur, BP 4229, 06304 Nice Cedex 4, France\\
$^{3}$Department of Space Studies, Southwest Research Institute, 1050 Walnut St., Suite 300, Boulder, CO 80302, USA}
\begin{document}

\date{Accepted ???. Received ???; in original form ???}

\pagerange{\pageref{firstpage}--\pageref{lastpage}} \pubyear{2010}

\maketitle

\label{firstpage}

%%%%%%%%%%%%%%%%%%%%%%%%%%%%%%%%%%%%%%%%%%%%%%%%%%%%%%%%%%%%%%%%%%%%%%%%

\begin{abstract}
We model the long-term evolution of the Hilda collisional family
located in the 3/2 mean-motion resonance with Jupiter.
Its eccentricity distribution evolves mostly due to the Yarkovsky/YORP effect
and assuming that:
(i) impact disruption was isotropic, and
(ii) albedo distribution of small asteroids is the same as for large ones,
we can estimate the age of the Hilda family to be $4_{-1}^{+0}\,{\rm Gyr}$.
We also calculate collisional activity in the J3/2 region.
Our results indicate that current collisional rates are very low
for a 200\,km parent body such that the number of expected events
over Gyrs is much smaller than one.

The large age and the low probability of the collisional disruption lead us
to the conclusion that the Hilda family might have been created
during the Late Heavy Bombardment when the collisions were much more frequent.
The Hilda family may thus serve as a test of orbital behavior of planets during the LHB.
We tested the influence of the giant-planet migration on the distribution of the family members.
The scenarios that are consistent with the observed Hilda family are those
with fast migration time scales $\simeq 0.3\,{\rm Myr}$ to $3\,{\rm Myr}$,
because longer time scales produce a family that is depleted and too much spread in eccentricity.
Moreover, there is an indication that Jupiter and Saturn were no longer in a compact configuration
(with period ratio $P_{\rm S}/P_{\rm J} > 2.09$) at the time when the Hilda family was created.

\end{abstract}

%%%%%%%%%%%%%%%%%%%%%%%%%%%%%%%%%%%%%%%%%%%%%%%%%%%%%%%%%%%%%%%%%%%%%%%%


%%%%%%%%%%%%%%%%%%%%%%%%%%%%%%%%%%%%%%%%%%%%%%%%%%%%%%%%%%%%%%%%%%%%%%%%

\section{Introduction}
There are many independent lines of evidence that the orbits of planets
of the Solar System were not the same all the time, but that they
have changed substantially over billions of years. The major arguments are based
on the observed orbital distribution of Kuiper belt objects (Malhotra et al. 1995, Levison et al. 2008)
or small but non-negligible eccentricities and inclinations of the giant planets (Tsiganis et al. 2005).
Observations of Jupiter's Trojans (Morbidelli et al. 2005),
main-belt asteroids (Minton \& Malhotra 2009, Morbidelli et al. 2010),
the amplitudes of secular oscillations of the planetary orbits (Morbidelli et al. 2009, Brasser et al. 2009),
or the existence of irregular moons (Nesvorn\'y et al. 2007)
provide important constraints for planetary migration scenarios.

Asteroids are a fundamental source of information about
the evolution of the planetary system.
Some of the resonant groups, i.e., located in the major mean-motion
resonances with Jupiter, might also have been influenced by planetary
migration, because their current distribution does not match
the map of the currently stable regions. For instance, there are two stable islands
denoted A and B in the J2/1 resonance and only the B~island is populated (Nesvorn\'y \& Ferraz-Mello 1997).

In this work we focus on the Hilda asteroid family in the 3/2 resonance with Jupiter.
We exploit our ability to model long-term evolution of asteroid families,
which is usually dominated by the Yarkovsky effect on the orbital elements (Bottke et al. 2001),
often coupled to the YORP effect on the spin rate and obliquity (Vokrouhlick\'y et al. 2006b).
Chaotic diffusion in eccentricity
and sometimes interactions with weak mean-motion or secular resonances (Vokrouhlick\'y et al. 2006a)
also play important roles.
In case of asteroids inside strong mean-motion resonances,
one has to account for the "resonant" {\em Yarkovsky effect\/},
which causes a systematic drift in eccentricity (Bro\v z \& Vokrouhlick\'y 2008).
This is different from usual non-resonant orbits where the Yarkovsky effect
causes a drift in semimajor axis.

The Hilda collisional family --- a part of the so called Hilda group in the 3/2 mean
motion resonance with Jupiter --- was already briefly discussed by Bro\v z \& Vokrouhlick\'y (2008).
However, the modelling presented in that paper was not very successful, since the resulting age of the family
seemed to be too large (exceeding $4\,{\rm Gyr}$). This was an important motivation for our current work.
We think that we missed an important mechanism in our previous model,
namely perturbations arising from the migration of the giant planets
and also an appropriate treatment of the YORP effect.
Indeed, the age $\gtrsim 4\,{\rm Gyr}$ suggests that the planetary migration might have played
a {\em direct\/} role during the early evolution of the Hilda family.
In this paper we thoroughly test this hypothesis.

The paper is organised as follows:
at first, we study the observed properties of the J3/2 resonance population in Section~\ref{sec:population}.
Our dynamical model of the Hilda family (without migration first) is described in Section~\ref{sec:model}.
Then we estimate the collisional activity in the J3/2 region in Section~\ref{sec:collisions}.
The results of our simulations of the giant-planet migration are presented in Section~\ref{sec:migration}.
Finally, Section~\ref{sec:conclusions} is devoted to conclusions.

%%%%%%%%%%%%%%%%%%%%%%%%%%%%%%%%%%%%%%%%%%%%%%%%%%%%%%%%%%%%%%%%%%%%%%%%

\section{Current asteroid population in the J3/2 resonance}\label{sec:population}

Asteroids located in the 3/2 mean motion resonance with Jupiter
have osculating semimajor axes around $(3.96\pm0.04)\,{\rm AU}$, i.e. beyond the main asteroid belt.
Contrary to the Kirkwood gaps (associated with J3/1, J7/3 or J2/1 resonances),
this resonance is populated by asteroids while its neighbourhood is almost empty.
The Hilda collisional family we are going to discuss in detail
is a small part of the whole J3/2 resonant population.

Our identification procedure of the J3/2 resonant population was described
in the previous paper Bro\v z \& Vokrouhlick\'y~(2008).
Using the AstOrb catalogue of orbits (version $JD = 2455500.5$, Oct 31st 2010)
we identified 1787 numbered and multi-opposition bodies with librating critical argument
\begin{equation}
\sigma = {p+q\over q} \lambda' - {p\over q} \lambda - \varpi\,,\label{sigma}
\end{equation}
where $p = 2$, $q = 1$,
$\lambda'$ is the mean longitude of Jupiter,
$\lambda$ the mean longitude of the asteroid
and $\varpi$ the longitude of perihelion of the asteroid.

In order to study the detailed distribution of the bodies librating inside the resonance
we have to use pseudo-proper resonant elements defined
as approximate surfaces of sections (Roig et al. 2002),
i.e. intersection of the trajectory with a plane defined by:
\begin{equation}
|\sigma| < 5^\circ\,,\quad
{\Delta\sigma\over\Delta t} > 0\,,\quad
|\varpi-\varpi'| < 5^\circ\,.\label{sigma_cond}
\end{equation}
These conditions correspond to the maximum of the semimajor axis $a$ over several oscillations
and the minimum of the eccentricity~$e$ or the inclination~$I$.
We need to apply a digital filter to $\sigma(t)$ prior to using Eq.~(\ref{sigma_cond}),
namely filter~A from Quinn, Tremaine \& Duncan (1991),
with sampling 1\,yr and decimation factor of 10,
to suppress fast $\simeq 80\,{\rm yr}$ oscillations, which would otherwise disturb
slower $\simeq 280\,{\rm yr}$ oscillations associated with resonant librations.
Finally, we apply an averaging of the sections $a, e, I$ over 1\,Myr running window
and these averages are the pseudo-proper elements $a_{\rm p}, e_{\rm p}, I_{\rm p}$.
The accuracy of the pseudo-proper elements is of the order $10^{-4}\,{\rm AU}$ for $a_{\rm p}$
and $10^{-4}$ for $e_{\rm p}$ or $\sin I_{\rm p}$ which is much smaller
than the structures we are interested in.

The overall dynamical structure of the J3/2 resonance is determined
by secular resonances $\nu_5$, $\nu_6$ at high eccentricities $e_{\rm p} \gtrsim 0.3$
and secondary resonances at lower values of $e_{\rm p} \lesssim 0.13$
(according to Morbidelli \& Moons 1993,
Nesvorn\'y \& Ferraz-Mello 1997,
Ferraz-Mello et al. 1998,
Roig \& Ferraz-Mello 1999).
They destabilise the orbits at the borders of a stable island.
The orbits inside the island exhibit very low chaotic diffusion rates,
so bodies can remain there for 4\,Gyr (without non-gravitational perturbation).

\begin{figure}
\centering
\includegraphics[width=8cm]{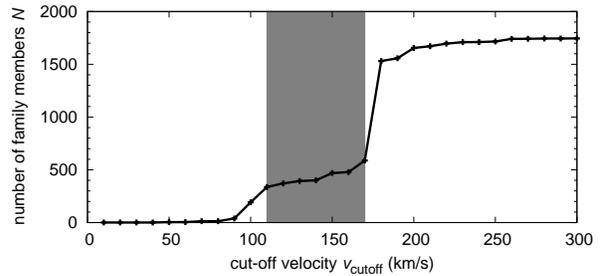}
\caption{The number $N$ of the Hilda family members versus the selected cut--off velocity $v_{\rm cutoff}$.}
\label{Nv}
\end{figure}

Next we apply a hierarchical clustering method (Zappal\`a et al. 1994)
to detect significant clusters. We use a standard metric in the pseudo-proper
element space $(a_{\rm p}, e_{\rm p}, \sin I_{\rm p})$
\begin{equation}
\delta v = na \sqrt{{5\over4} \left({\delta a_{\rm p}\over a_{\rm p}}\right)^2 + 2 (\delta e_{\rm p}^2) + 2 (\delta \sin I_{\rm p})^2}\,.\label{eq:delta_v}
\end{equation}
In the following, we do not discuss the known Schubart family, which was sufficiently
analysed elsewhere (Bro\v z \& Vokrouhlick\'y 2008), but we focus on the family associated with (153)~Hilda.
A suitable cut--off velocity for the Hilda family
seems to be $v_{\rm cutoff} = 140\,{\rm m}/{\rm s}$,
because the number of members does not change substantially
around this value (see Figure~\ref{Nv}).
The number of members at this cut--off is~400.

\begin{figure*}
\centering
\includegraphics[width=58mm]{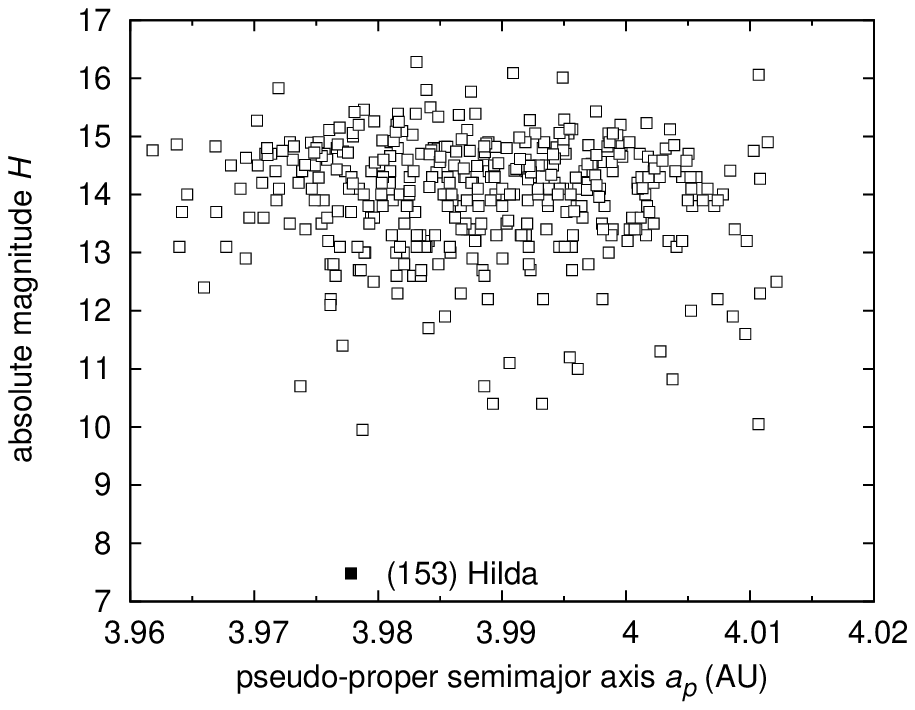}%
\includegraphics[width=58mm]{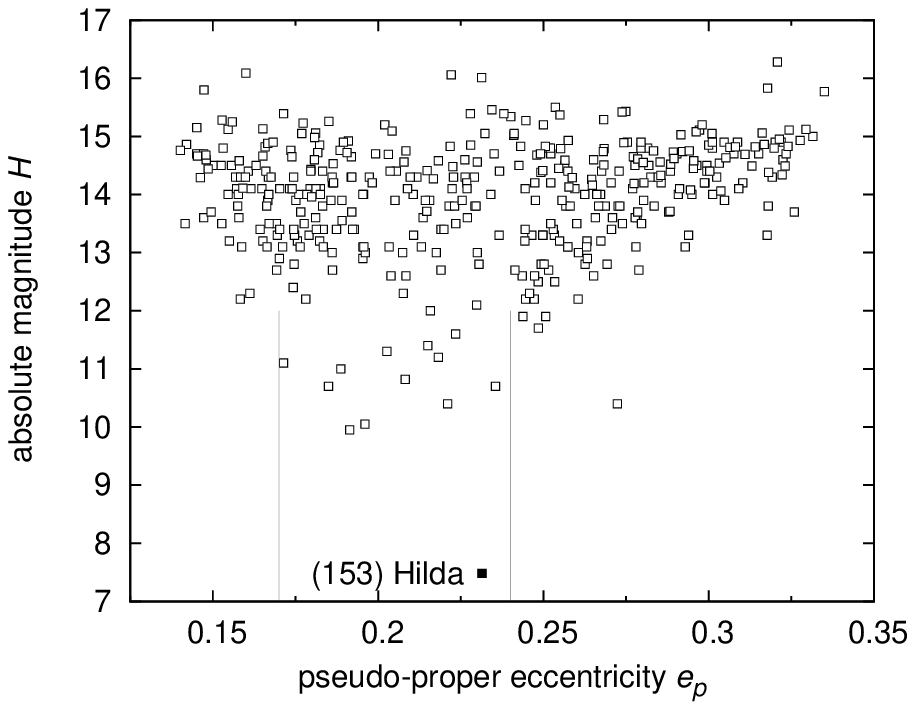}%
\includegraphics[width=58mm]{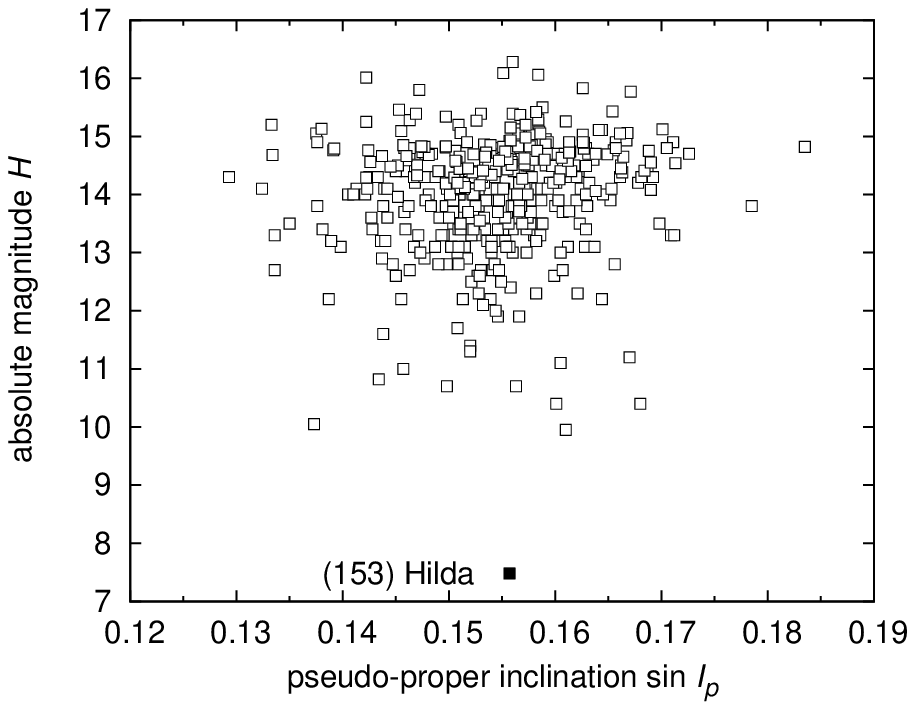}
\caption{The Hilda family displayed in
resonant semimajor axis $a_{\rm p}$ (left), eccentricity $e_{\rm p}$ (middle) and inclination $\sin I_{\rm p}$ (right) versus absolute magnitude~$H$.
The libration centre is located at $a \simeq 3.96\,{\rm AU}$ and all bodies are displayed to the right of it.
The `ears' in $(e_{\rm p},H)$, i.e., the concentration of small asteroids at the outskirts
of the family and their depletion in the centre, are very prominent here.
The thin vertical lines denote the central part of the $(e_{\rm p}, H)$ distribution discussed in the text.
The family has 400~members at $v_{\rm cutoff} = 140\,{\rm m}/{\rm s}$.}
\label{hilda_aHeHiH}
\end{figure*}

The resulting plots $(a_{\rm p},H)$, $(e_{\rm p},H)$ and $(I_{\rm p},H)$ of the Hilda family
show very interesting features (see Figure~\ref{hilda_aHeHiH}).
The distribution of semimajor axis and inclination seems rather uniform and almost independent
of absolute magnitude~$H$,
but eccentricities of small asteroids (i.e., with high~$H$)
are clearly concentrated at the outskirts of the family and depleted in the centre.

In order to explain the distribution of asteroids in the $(e_{\rm p}, H)$ plane
we have to recall that asteroids orbiting about the Sun are affected by non-gravitational forces,
mostly by the Yarkovsky/YORP effect, i.e. the recoil force/torque due
to anisotropic emission of thermal radiation.
We consider the concentrations in the $(e_{\rm p}, H)$ plane to be a strong indication
of the ongoing Yarkovsky/YORP evolution, because they are very similar to those
observed among several main-belt families in the $(a_{\rm p}, H)$ plane
and successfully modelled by Vokrouhlick\'y et al. (2006b). The difference
between these two cases stems from the fact that main-belt families are non-resonant
and the Yarkovsky/YORP effect thus increases or decreases the semimajor axis
(depending on the actual obliquity of the spin axis), while in our resonant case,
the same perturbation results instead in a systematic increase or decrease of eccentricity.
A detailed modelling of the $e$-distribution is postponed to Section~\ref{sec:results_yarko}.

\begin{figure}
\centering
\includegraphics[width=68mm]{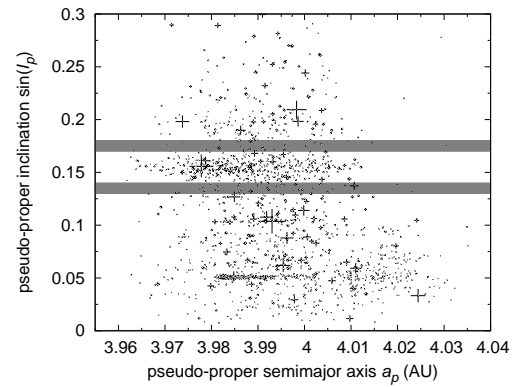}
\caption{The J3/2 region displayed in $(a_{\rm p}, I_{\rm p})$ plot.
A very prominent Schubart cluster (studied by Bro\v z \& Vokrouhlick\'y 2008)
is visible around $\sin I_{\rm p} \doteq 0.05$.
The close surroundings of the Hilda family, where only a low number of bodies is present,
is highlighted by grey rectangles.}
\label{r32_EVEN2_ai_4000Myr}
\end{figure}

The central part of the $(e_{\rm p}, H)$ distribution, from $e = 0.17$ to 0.23,
seems rather extended. The large asteroids ($H < 12.5\,{\rm mag}$) are spread
over this interval of eccentricities even though their Yarkovsky drift rates must have been small.
Only 2--4 of them are likely to be interlopers, because there is a very low number of background asteroids
in the surroundings of the family (see Figure~\ref{r32_EVEN2_ai_4000Myr}).
We think this shape might actually be the result of the initial size-independent
perturbation that the family distribution received by the migration of the giant planets
(which we discuss in Section~\ref{sec:results_mig}).

Regarding the $(a_{\rm p}, H)$ distribution, the largest asteroid (153) Hilda
is offset with respect to the centre, but this is a natural outcome of the
definition of the pseudo-proper elements --- fragments which fall to the left
of the libration centre are mapped to the right which creates the offset.

The geometric albedos for Hilda family objects are poorly known.
There are only six measured values for the family members:
0.064,
0.046,
0.038,
0.089,
0.044,
0.051
(Davis \& Neese 2002).
Given the low number of values and the possibility of selection effects
we prefer to assume the family members have a mean value $p_V = 0.044$,
which corresponds to the whole J3/2 population.
The size of the parent body can be then estimated to be $D_{\rm PB} = (200\pm20)\,{\rm km}$.
We employ two independent methods to determine the diameter $D_{\rm PB}$:
 (i)~we sum the volumes of the observed bodies larger than an assumed completeness limit
     $D_{\rm complete} = 10\,{\rm km}$ and then we prolong the slope of the size-frequency
     distribution down to $D = 0$ to account for unobservable bodies (see Bro\v z \& Vokrouhlick\'y 2008),
     which results in $D_{\rm PB} \simeq 185\,{\rm km}$;
(ii)~we also use a geometric method developed by Tanga et al. (1998)
which gives $D_{\rm PB} \simeq 210\,{\rm km}$.
A test with different albedo values will be described in Section~\ref{sec:high_albedo}.

\begin{figure}
\centering
\includegraphics[width=58mm]{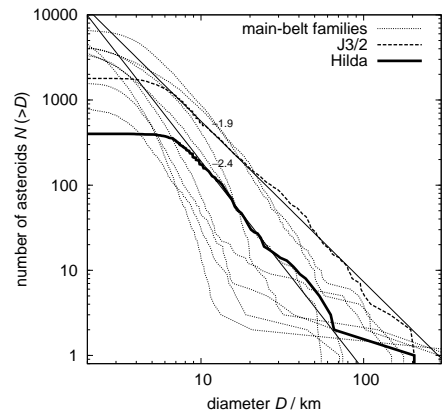}
\caption{Cumulative size distributions of the J3/2 population and the Hilda family.
The polynomial fits of the form $N({>}D) = C D^\gamma$ are plotted as thin lines,
together with the respective values of the $\gamma$ exponent.
Several main-belt families are plotted for comparison:
Eos (with slope $\gamma = -2.8$),
Eunomia   ($-5.0$),
Hygiea    ($-3.8$),
Koronis   ($-2.8$),
Themis    ($-2.9$),
Tirela    ($-3.3$),
Veritas   ($-3.4$)
and Vesta ($-5.4$).}
\label{r32-size_distribution_hilda_INC3}
\end{figure}

The size-frequency distribution $N({>}D)$ vs $D$ of the Hilda family is steeper
than that of background J3/2 population, but shallower than for usual main-belt families
(Figure~\ref{r32-size_distribution_hilda_INC3}). Interestingly, the slope $\gamma = -2.4\pm 0.1$
of the distribution $N({>}D) = C D^\gamma$ is close to a collisional equilibrium
calculated by Dohnanyi (1969).

Colour data extracted from the Sloan Digital Sky Survey Moving Object Catalogue version~4
(Parker et al. 2008) confirm the Hilda family belongs to the taxonomic type~C,
because most of the spectral slopes are small. Recall that the whole J3/2 population
exhibits a bimodal distribution of slopes, i.e. it contains a mixture
of C- and D-type asteroids.

%%%%%%%%%%%%%%%%%%%%%%%%%%%%%%%%%%%%%%%%%%%%%%%%%%%%%%%%%%%%%%%%%%%%%%%%

\section{The Hilda family model with radiation forces}\label{sec:model}

To understand the long term evolution of the Hilda family, we construct
a detailed numerical model, extending efforts in Bro\v z \& Vokrouhlick\'y (2008),
which includes the following processes:
  (i)~impact disruption,
 (ii)~the Yarkovsky effect,
(iii)~the YORP effect,
 (iv)~collisions and spin-axis reorientations.
We describe the individual parts of the model in the forthcoming subsections.

\subsection{Impact disruption}

To obtain initial conditions for the family just after the breakup event
we need a model for the ejection velocities of the fragments.
We use a very simple model of an isotropic ejection from
the work of Farinella et al. (1994). The distribution of velocities
"at infinity" follows the function
\begin{equation}
{\rm d}N(v) {\rm d} v = C v (v^2 + v_{\rm esc}^2)^{-(\alpha+1)/2} {\rm d} v\,,\label{dN_v}
\end{equation}
with the exponent $\alpha$ being a~free parameter,
$C$ a normalisation constant
and $v_{\rm esc}$ the escape velocity from the parent body,
which is determined by its size~$D_{\rm PB}$ and mean density~$\rho_{\rm PB}$ as
$v_{\rm esc} = \sqrt{(2/3) \pi G \rho_{\rm PB}}\, D_{\rm PB}\,.$
The distribution is usually cut at a selected maximum allowed 
velocity $v_{\rm max}$ to prevent outliers.
The actual values of all these parameters are given in Section~\ref{sec:results_yarko}.
Typically, the overall distribution of velocities has a peak close to the escape velocity,
which is approximately 100\,m/s for a 200\,{\rm km} parent body.
The initial velocities $|v|$ of individual bodies are generated by a straightforward Monte--Carlo code
and the orientations of the velocity vectors $\vec v$ in space are assigned randomly.

Here, we assume the velocity of fragments is independent of their size,
which seems reasonable with respect to the observed uniform distribution
of the Hilda family in the $(a_{\rm p}, H)$ and $(I_{\rm p}, H)$ planes (Figure~\ref{hilda_aHeHiH}).
We perform also tests with non-isotropic distributions in Section~\ref{sec:asymmetric}.

We must also select initial osculating eccentricity~$e_{\rm i}$ of the parent body,
initial inclination~$i_{\rm i}$,
as well as true anomaly~$f_{\rm imp}$
and argument of perihelion~$\omega_{\rm imp}$
at the time of impact disruption.
All of these parameters determine the initial shape of the synthetic
"Hilda" family just after the disruption of the parent body.
Initial semimajor axis~$a_{\rm i}$ is not totally free, instead it is
calculated from the initial semimajor axis of Jupiter~$a_{\rm Ji}$
and the Kepler law, since the parent body has to be confined in the J3/2 resonance.

%%%%%%%%%%%%%%%%%%%%%%%%%%%%%%%%%%%%%%%%%%%%%%%%%%%%%%%%%%%%%%%%%%%%%%%%

\subsection{Yarkovsky effect in a resonance}

The long-term evolution of asteroid orbits is mainly driven by the Yarkovsky thermal effect.
The implementation of the Yarkovsky effect in the SWIFT integrator
was described in detail in Bro\v z (2006). Only minor modifications of the
code were necessary to incorporate spin rate evolution, which is driven
by the YORP effect (see Section~\ref{sec:yorp}).

The thermal parameter we use are reasonable estimates for C/X-type bodies:
$\rho_{\rm surf} = \rho_{\rm bulk} = 1300\,{\rm kg}/{\rm m}^3$ for the surface and bulk densities,
$K = 0.01\,{\rm W}/{\rm m}/{\rm K}$ for the surface thermal conductivity,
$C = 680\,{\rm J}/{\rm kg}$ for the heat capacity,
$A = 0.02$ for the Bond albedo and
$\epsilon_{\rm IR} = 0.95$ for the thermal emissivity parameter.

\begin{figure}
\centering
\includegraphics[width=7cm]{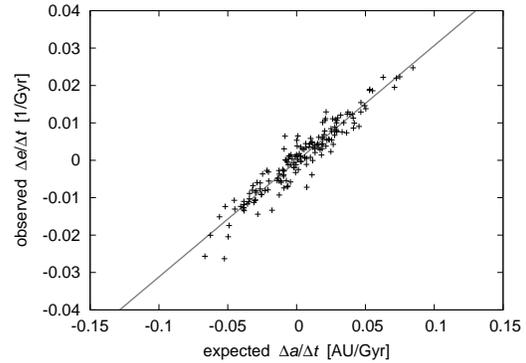}
\caption{Almost linear relation between the expected drift $\Delta a$ in semimajor axis
and the simulated drift $\Delta e$ in eccentricity, computed for 360 members of the Hilda family
located inside the J3/2 resonance.}
\label{hilda_deltae}
\end{figure}

We can use a standard algorithm for the calculation of the Yarkovsky acceleration
which results in a semimajor-axis drift in case of non-resonant bodies.
The drift in eccentricity in case of resonant bodies arises 
"automatically" due to the gravitational part of the integrator.
In Figure~\ref{hilda_deltae} we can see a comparison between the
expected drift~$\Delta a$ in semimajor axis and the resulting drift~$\Delta e$
in eccentricity, computed for the Hilda family
(see the explanation in Appendix A of Bro\v z \& Vokrouhlick\'y 2008).
The data can be approximated by a linear relationship, where the departures from linearity
are caused mainly by interactions of drifting orbits with embedded weak secular
or secondary resonances.

Note that according to a standard solar model the young Sun was faint
(G\"udel 2007), i.e., its luminosity 4\,Gyr ago was 75\,\% of the current~$L_\odot$.
We can then expect a lower insolation and consequently weaker thermal effects acting on asteroids.
Since we assume a constant value of~$L_\odot$ in our code
the age estimated for the Hilda family (in Section~\ref{sec:results_yarko}) can be 12.5\,\% larger.

%%%%%%%%%%%%%%%%%%%%%%%%%%%%%%%%%%%%%%%%%%%%%%%%%%%%%%%%%%%%%%%%%%%%%%%%

\subsection{YORP effect}\label{sec:yorp}

The magnitude of the Yarkovsky drift sensitively depends on the orientation
of the spin axis with respect to the orbital plane and, to a lesser extent, on the angular velocity too.
We thus have to account for the long-term evolution of spins of asteroids
which is controlled by torques arising from the emission of thermal radiation, i.e. the YORP effect.
The implementation of the YORP effect follows Vokrouhlick\'y et al. (2006).
We assume the following relations for the rate of angular velocity
and obliquity
\begin{eqnarray}
{{\rm d}\omega\over{\rm d} t} &=& f_i(\epsilon)\,,\qquad i = 1 \dots 200\,,\label{eq:domega}\\
{{\rm d}\epsilon\over{\rm d} t} &=& {g_i(\epsilon)\over\omega}\,,\label{eq:depsil}
\end{eqnarray}
where $f$- and $g$-functions are given by \v Capek \& Vokrouhlick\'y (2004)
for a set of 200 shapes with
mean radius $R_0 = 1\,{\rm km}$,
bulk density $\rho_0 = 2500\,{\rm kg}/{\rm m}^3$,
located on a circular orbit with semimajor axis $a_0 = 2.5\,{\rm AU}$.
The shapes of the Hilda family members are not known,
so we assign one of the artificial shapes (denoted by the index~$i$) randomly to each individual asteroid.
We only have to scale the $f$- and $g$-functions by a factor
\begin{equation}
c = c_{\rm YORP} \left({a\over a_0}\right)^{-2} \left(R\over R_0\right)^{-2} \left(\rho_{\rm bulk}\over\rho_0\right)^{-1}\,,
\end{equation}
where $a$, $R$, $\rho_{\rm bulk}$ are semimajor axis, radius and density of the simulated body,
and $c_{\rm YORP}$ is a free scaling parameter, which can account for an additional uncertainty of the YORP model.
Because the values of $f$'s and $g$'s were computed for only a limited set of obliquities (with a~step $\Delta\epsilon = 30^\circ$)
we use interpolation by Hermite polynomials (Hill 1982) of the data in \v Capek \& Vokrouhlick\'y (2004)
to obtain a smooth analytical functions for $f_i(\epsilon)$ and $g_i(\epsilon)$.

If the angular velocity approaches a critical value
\begin{equation}
\omega_{\rm crit} = \sqrt{{8\over 3} \pi G \rho_{\rm bulk}}\,,
\end{equation}
we assume a mass shedding event, so we keep the orientation of
the spin axis and the sense of rotation, but we reset the orbital period~$P = {2\pi/\omega}$
to a random value from the interval $(2.5, 9)$~hours.
We also change the assigned shape to a different one,
since any change of shape may result in a different YORP effect.
% We do not change the mass, however.

\begin{figure}
\includegraphics[height=4.3cm]{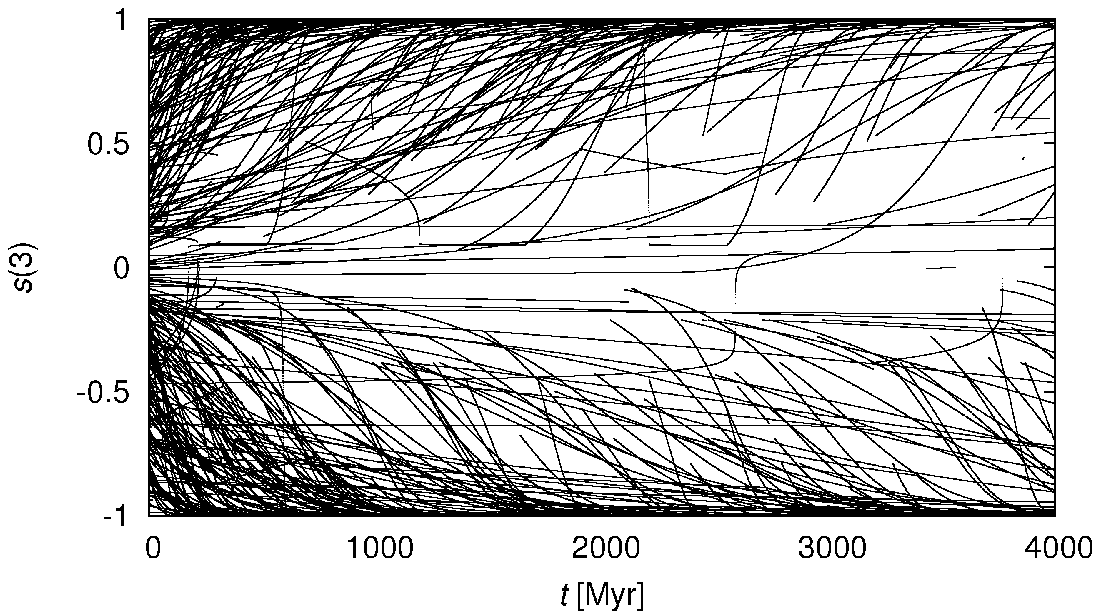}
\includegraphics[height=4.3cm]{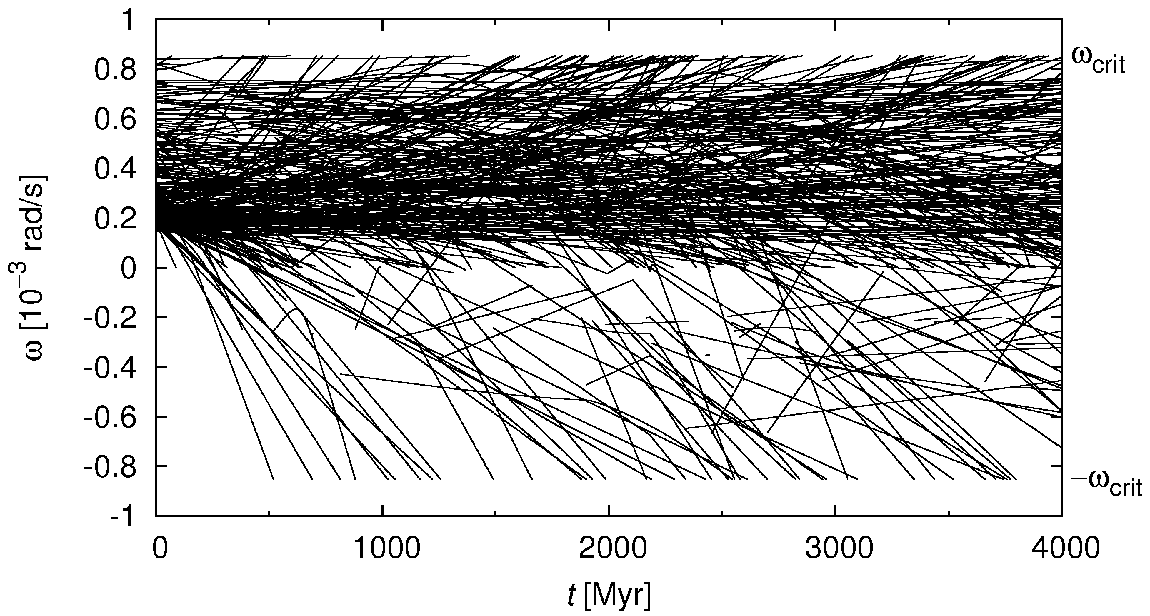}
\caption{An example of the YORP-driven evolution of obliquities (namely a $z$-component of the spin axis unit vector, top panel)
and angular velocities~$\omega$ (bottom panel) for the members of the synthetic "Hilda" family.
At the beginning, all values of $\omega$ were selected positive and spin axes were distributed isotropically.
The evolution may force $\omega$ to become negative, which simply corresponds to an opposite orientation of the spin axis.
The scaling parameter was selected $c_{\rm YORP} = 0.33$ in this run.}
\label{r32-impye3_YORP_cYORP_0.33_reorient}
\end{figure}

The differential equations~(\ref{eq:domega}), (\ref{eq:depsil})
are integrated numerically by a simple Euler integrator.
The usual time step is $\Delta t = 1000\,{\rm yr}$.
An example of the results computed by the spin integrator for the Hilda family
is displayed in Figure~\ref{r32-impye3_YORP_cYORP_0.33_reorient}.
The typical time scale of the spin axis evolution is $\tau_{\rm YORP} \simeq 500\,{\rm Myr}$.
After $\simeq3$ times $\tau_{\rm YORP}$ most bodies have spin axes perpendicular
to their orbits, what maximizes the Yarkovsky drift rate of eccentricity.

%%%%%%%%%%%%%%%%%%%%%%%%%%%%%%%%%%%%%%%%%%%%%%%%%%%%%%%%%%%%%%%%%%%%%%%%

\subsection{Collisions and spin-axis reorientations}

In principle, collisions may directly affect the size distribution of the synthetic "Hilda" family,
but we neglect this effect because most of the asteroids are large enough to remain intact.

However, we include spin axis reorientations caused by collisions.
We use an estimate of the time scale by Farinella et al. (1998)
\begin{equation}
\tau_{\rm reor} = B \left({\omega\over\omega_0}\right)^{\beta_1} \left({D \over D_0}\right)^{\beta_2}\,,
\end{equation}
where
$B = 84.5\,{\rm kyr}$,
$\beta_1 = 5/6$,
$\beta_2 = 4/3$,
$D_0 = 2\,{\rm m}$ and
$\omega_0$ corresponds to period $P = 5$~hours.
These values are characteristic for the main belt and
we use them as an upper limit of $\tau_{\rm reor}$ for the J3/2 region.
Even so, the time scale is $\tau_{\rm reor} \simeq 3\,{\rm Gyr}$
for the smallest observable ($D \simeq 5\,{\rm km}$) bodies
and reorientations are thus only of minor importance.
Note that the probability of the reorientation is enhanced
when the YORP effect drives the angular velocity~$\omega$ close to zero.

%%%%%%%%%%%%%%%%%%%%%%%%%%%%%%%%%%%%%%%%%%%%%%%%%%%%%%%%%%%%%%%%%%%%%%%%

\subsection{Results on Yarkovsky/YORP evolution}\label{sec:results_yarko}

\begin{figure}
\centering
\includegraphics[width=6.5cm]{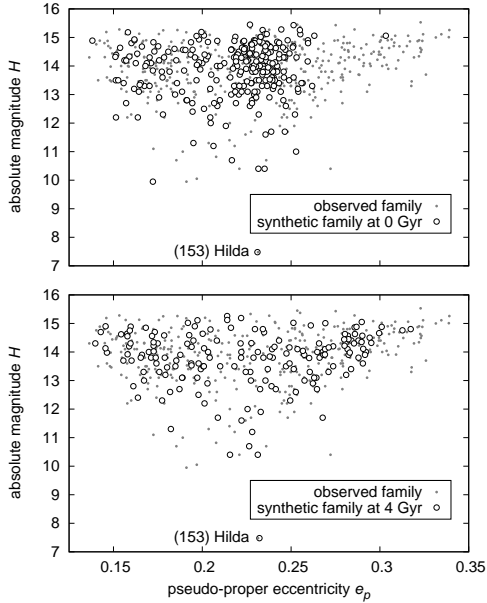}
\caption{Eccentricity vs absolute magnitude plot for the synthetic "Hilda" family
just after the impact disruption (time $t = 0$, top panel) and after 4\,Gyr of evolution
due to the Yarkovsky/YORP effect (bottom panel).
There is a comparison with the observed Hilda family (gray dots).}
\label{r32-impye3_YORP_cYORP_0.33_hilda_eH_4000Myr}
\end{figure}

We start a simulation with an impact disruption of the parent body and create 360~fragments.
Subsequent evolution of the synthetic "Hilda" family due to the Yarkovsky/YORP effect
is computed up to $6\,{\rm Gyr}$ in order to estimate the time span needed to match the observed
family even though the family cannot be older than $\simeq 4\,{\rm Gyr}$, of course.
Planets are started on their current orbits. A typical outcome of the simulation is displayed
in Figure~\ref{r32-impye3_YORP_cYORP_0.33_hilda_eH_4000Myr}.

Due to the long integration time span and large number of bodies,
we were able to compute only four simulations with the following values
of true anomaly and YORP efficiency:
\begin{enumerate}
\item $f_{\rm imp} = 0^\circ$, $c_{\rm YORP} = 0$;
\item $f_{\rm imp} = 180^\circ$, $c_{\rm YORP} = 0$;
\item $f_{\rm imp} = 0^\circ$, $c_{\rm YORP} = 1$;
\item $f_{\rm imp} = 0^\circ$, $c_{\rm YORP} = 0.33$.
\end{enumerate}
The remaining parameters were fixed:
$e_{\rm i} = 0.14$,
$i_{\rm i} = 7.8^\circ$,
$\omega_{\rm imp} = 30^\circ$,
$\alpha = 3.25$,
$v_{\rm max} = 300\,{\rm m}/{\rm s}$,
$R_{\rm PB} = 93.5\,{\rm km}$,
$\rho_{\rm PB} = 1300\,{\rm kg}/{\rm m}^3$,
$p_V = 0.044$.

\begin{figure}
\centering
\includegraphics[width=6cm]{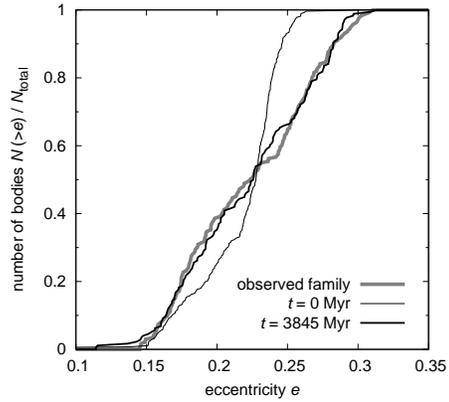}
\caption{Normalized cumulative distributions $N({<}e)$ of eccentricities for
  (i)~the observed Hilda family,
 (ii)~the synthetic "Hilda" family at time $t = 0$ (just after the impact disruption),
(iii)~evolved due to the Yarkovsky/YORP effect (at time $t = 3845\,{\rm Myr}$).
In this figure we show the best fit for the simulation with parameters $f_{\rm imp} = 0^\circ$, $c_{\rm YORP} = 0.33$.
Note the `bended' shape of the observed distribution corresponds to the `ears' on the $(e_{\rm p}, H)$ plot
(Figure~\ref{hilda_aHeHiH}). There is no perturbation by planetary migration in this particular case.}
\label{r32-impye3_YORP_cYORP_0.33}
\end{figure}

\begin{figure*}
\centering
\begin{tabular}{cccc}
(a)~Yarkovsky effect only, $f_{\rm imp} = 0^\circ$ &
(b)~$f_{\rm imp} = 180^\circ$ \\
\includegraphics[width=7cm]{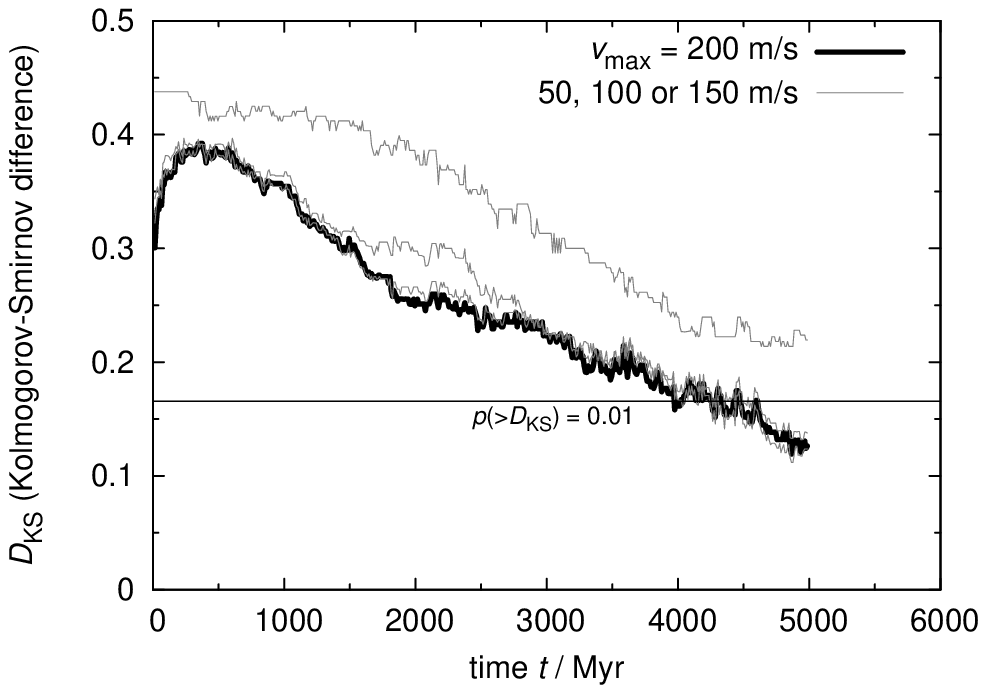} &
\includegraphics[width=7cm]{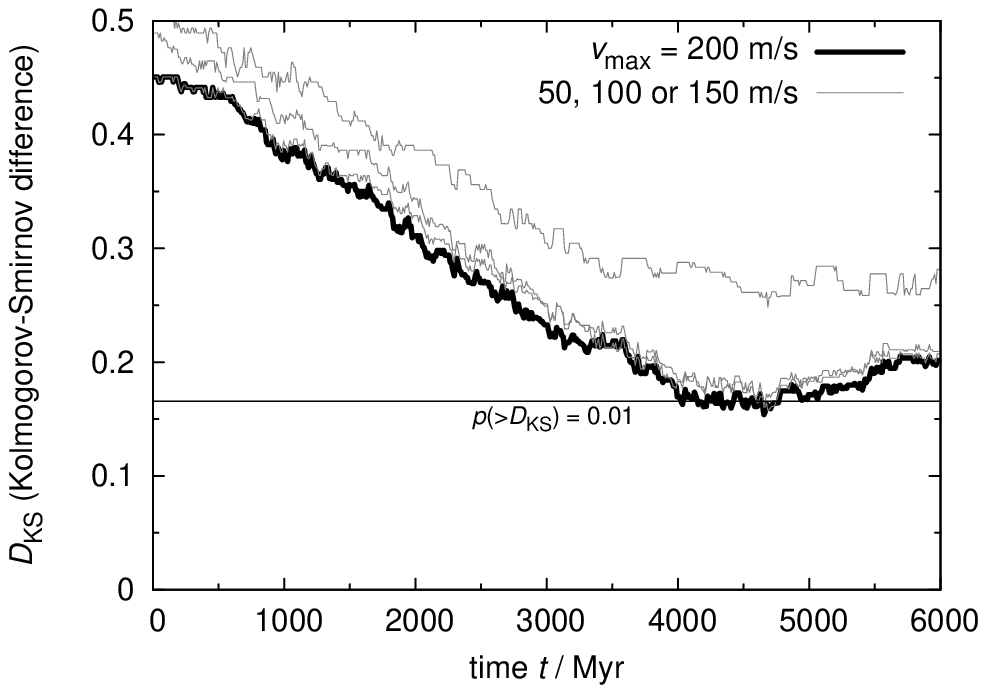} \\
(c)~YE with YORP effect, $c_{\rm YORP} = 1$ &
(d)~YORP, $c_{\rm YORP} = 0.33$ \\
\includegraphics[width=7cm]{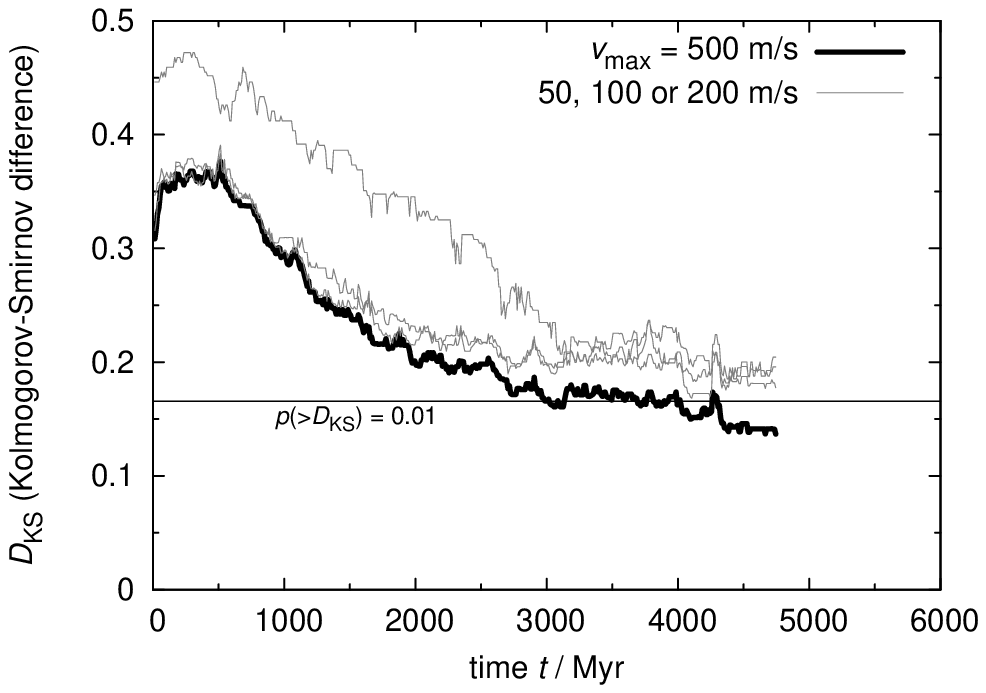} &
\includegraphics[width=7cm]{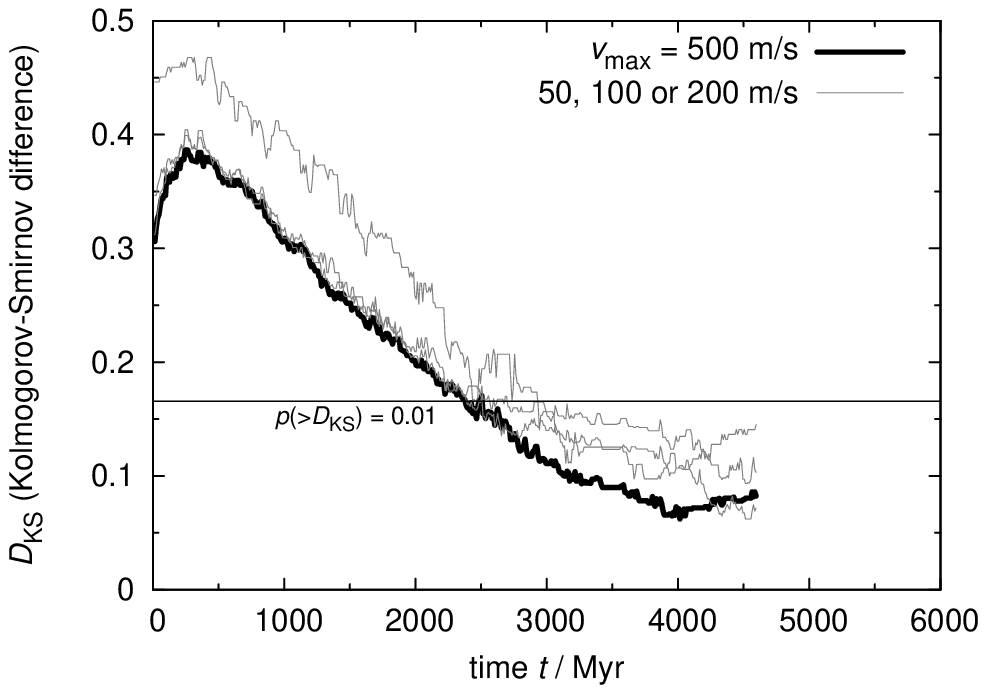} \\
(e)~high albedo &
(f)~asymmetric velocity field \\
\includegraphics[width=7cm]{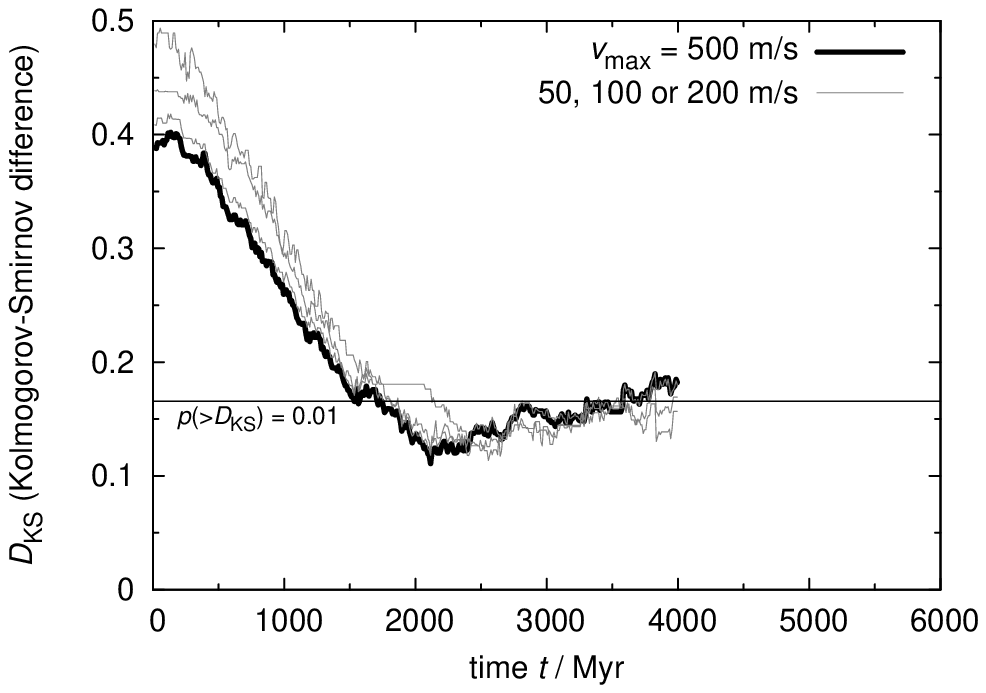} &
\includegraphics[width=7cm]{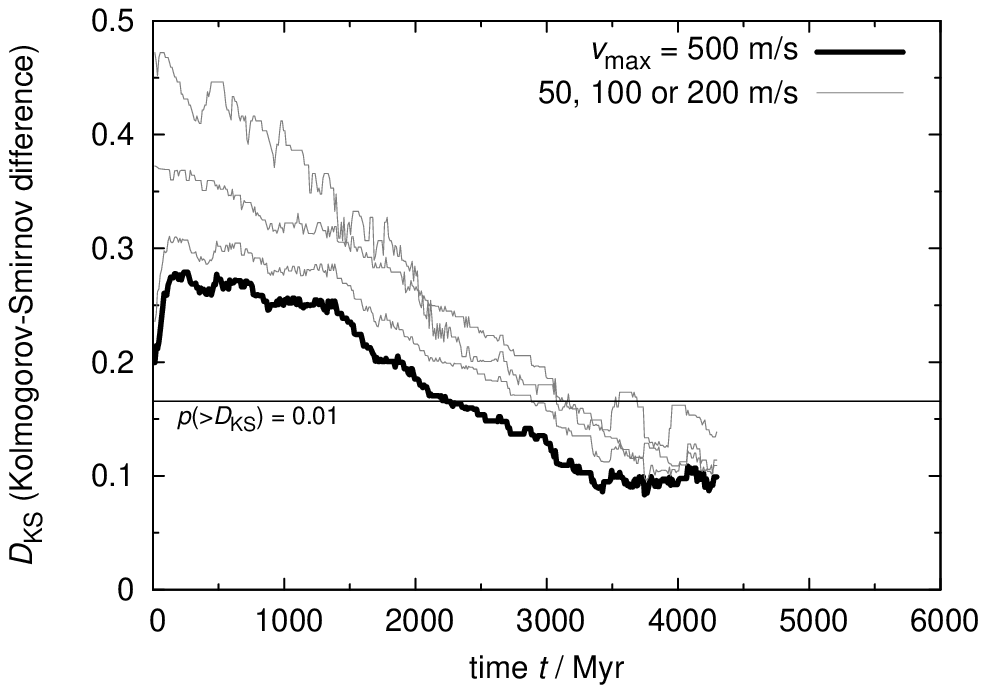} \\
\end{tabular}
\caption{Kolmogorov-Smirnov tests of the synthetic "Hilda" family:
(a)~no migration, only initial disruption (at anomaly $f_{\rm imp} = 0^\circ$, $\varpi_{\rm i} = 30^\circ$) and subsequent Yarkovsky evolution;
(b)~the case with $f_{\rm imp} = 180^\circ$;
(c)~including the YORP effect;
(d)~YORP with efficiency factor $c_{\rm YORP} = 0.33$;
(e)~high albedo values (i.e., small bodies);
(f)~strongly asymmetric velocity field.
The horizontal line denotes the distance $D_{\rm KS} = 0.165$ for which the probability
$p({>}D_{\rm KS})$ that the two eccentricity distributions differ by this amount equals to 0.01.}
\label{r32-impye3_hilda_kstest1_D_VTEST}
\label{r32-impye3_f180_hilda_kstest1_D_VTEST}
\label{r32-impye3_PUSH_LIMITS_hilda_kstest1_D_VTEST}
\label{r32-impye3_ASYMMETRIC_VFIELD_hilda_kstest1_D_VTEST}
\label{r32-impye3_YORP_hilda_kstest1_D_VTEST}
\label{r32-impye3_YORP_cYORP_0.33_hilda_kstest1_D_VTEST}
\end{figure*}

We are mainly concerned with the distribution of eccentricities~$e_{\rm p}$,
because the observed family has a large spread of $e_{\rm p}$'s,
while the initial synthetic family is very compact. For this purpose we constructed
a Kolmogorov--Smirnov test (Press et al. 1999) of the normalized cumulative distributions $N({<}e)$
\begin{equation}
D_{\rm KS} = \max_{0 < e < 1} |N({<}e)_{\rm syn} - N({<}e)_{\rm obs}|\,,
\end{equation}
which provides a measure of the difference between the synthetic "Hilda" family,
at a given time, and the observed Hilda family (see Figure~\ref{r32-impye3_YORP_cYORP_0.33} for an example).
The results of the KS tests are summarized in Figure~\ref{r32-impye3_hilda_kstest1_D_VTEST} (first four panels).

There is an easy possibility to asses the sensitivity of results with respect to
the $v_{\rm max}$ parameter too, without the need to compute the simulation again.
We simply select bodies fulfilling the condition $v < v_{\rm max}'$,
with $v_{\rm max}' = 200$, 100 or $50\,{\rm m}/{\rm s}$,
and recompute only the KS statistics for this subset. The results are plotted
in Figure~\ref{r32-impye3_hilda_kstest1_D_VTEST} as thin lines.
We can state values lower than $v_{\rm max} \simeq 100\,{\rm m}/{\rm s}$
are surely excluded.

As a preliminary conclusion we may say that all simulations point
to a large age of the Hilda family. The $e_{\rm p}$-distributions are
most compatible with the observed family for ages $t = (4.0\pm 1.0)\,{\rm Gyr}$.
This suggests the Hilda family might have experienced the giant-planet migration period
which is dated by the Late Heavy Bombardment to $t_{\rm LHB} \simeq 3.85\,{\rm Gyr}$
(Gomes et al. 2005).
The large uncertainty of the age stems from the fact that the runs including the YORP effect
($c_{\rm YORP} \ge 0.33$) tend to produce ages at a lower limit of the interval
while the YORP-less runs (with $c_{\rm YORP} = 0$) tend to the upper limit.

%%%%%%%%%%%%%%%%%%%%%%%%%%%%%%%%%%%%%%%%%%%%%%%%%%%%%%%%%%%%%%%%%%%%%%%%

\subsection{Alternative hypothesis: high albedos of small asteroids}\label{sec:high_albedo}

We now discuss two scenarios that further reduce the minimal age of the family:
 (i)~high albedos of small asteroids (i.e., larger Yarkovsky/YORP drift);
(ii)~strongly asymmetric velocity field after impact (like that of the Veritas family).

Albedo is the most important unknown parameter, which can affect results
on the Yarkovsky/YORP evolution. Fern\'andez et al. (2009) measured
albedos of small Trojan asteroids and found a systematically larger
values than for large Trojans. If we assume the J3/2 asteroids behave
similarly to Trojans, we may try a simulation with an rather high
value of geometric albedo $p_V = 0.089$ (compared to previous $p_V = 0.044$).
Moreover, we decrease density $\rho_{\rm bulk} = 1200\,{\rm kg}/{\rm m}^3$,
increase maximum velocity of fragments $v_{\rm max} = 500\,{\rm m}/{\rm s}$
(though the velocity distribution is still determined by Eq.~\ref{dN_v})
and select true anomaly $f_{\rm imp} = 90^\circ$
to maximise the spread of~$e_{\rm p}$'s.

The KS test is included in Figure~\ref{r32-impye3_PUSH_LIMITS_hilda_kstest1_D_VTEST}, panel~(e).
The most probable age is $(2.3\pm0.5)\,{\rm Gyr}$ in this case.
However, we do not think that the size-dependent albedo is very plausible
because both large and small family members
should originate from the same parent body and their albedos, at least just after
the disruption, should be similar. Nevertheless, the albedos may change
to a certain degree due to space weathering processes (Nesvorn\'y et al. 2005).
Unfortunately, we do not have enough data for small asteroids to assess
a possible albedo difference between large and small family members.

%%%%%%%%%%%%%%%%%%%%%%%%%%%%%%%%%%%%%%%%%%%%%%%%%%%%%%%%%%%%%%%%%%%%%%%%

\subsection{Alternative hypothesis: strongly asymmetric velocity field}\label{sec:asymmetric}

Another possibility to reduce estimate of the family age is that the original velocity was highly anisotropic.
A well known example from the main belt is the Veritas family.
Let us assume the anisotropy is of the order of Veritas, i.e.,
approximately 4 times larger in one direction.
Note that Veritas is a young family and can be modelled precisely enough
to compensate for chaotic diffusion in resonances (Nesvorn\'y et al. 2003, Tsiganis et al. 2007).
This family is characteristic by a large spread in inclinations,
which corresponds to large out-of-plane components of velocities.
In case of the Hilda family we multiply by~4 the radial components of initial velocities
to maximise the dispersion of eccentricities, assuming the most favourable geometry of disruption ($f_{\rm imp} \doteq 90^\circ$).

The fit in Figure~\ref{r32-impye3_ASYMMETRIC_VFIELD_hilda_kstest1_D_VTEST}, panel~(f)
is seemingly better at the beginning of the simulation,
but bodies on unstable orbits are quickly eliminated and the fit
gets much worse at $t \simeq 500\,{\rm Myr}$. We can see that the synthetic
"Hilda" family is similar to the observed Hilda family quite early
(at $t \simeq 2.5\,{\rm Gyr}$), however the best fit is at later times
($t \simeq 3.5\,{\rm Gyr}$), so there is no significant benefit
compared to isotropic velocity-distribution cases.

%%%%%%%%%%%%%%%%%%%%%%%%%%%%%%%%%%%%%%%%%%%%%%%%%%%%%%%%%%%%%%%%%%%%%%

\section{Disruption rates in the J3/2 population}\label{sec:collisions}

\subsection{Present collisional activity}\label{sec:collisions_present}

The results presented above show that the Hilda family is old.
However, the uncertainty of the age is too large to conclude
whether the family formed during the LHB period.
An alternative constraint is the collisional lifetime of the parent body.
If the probability that the parent body broke in the last 4\,Gyr
in the current collisional environment is negligible,
this would argue that the family broke during the LHB
when the collisional bombardment was much more severe.
Thus, here we estimate the collisional lifetime of the parent body.

In our case, the target (parent body) has diameter $D_{\rm target} = 200\,{\rm km}$,
mean impact velocity $V_{\rm i} = 4.8\,{\rm km}/{\rm s}$ (Dahlgren 1998),
probable strength $Q^\star_D = 4\times 10^5\,{\rm J}/{\rm kg}$ (Benz \& Asphaug 1999)
and thus the necessary impactor size (Bottke et al. 2005) is
\begin{equation}
d_{\rm disrupt} = \left({2 Q^\star_D / V_{\rm i}^2}\right)^{1/3} D_{\rm target}
\simeq 65\,{\rm km}\,.\label{eq:d_disrupt}
\end{equation}
The population of ${\ge}65\,{\rm km}$ projectiles is dominated by main-belt bodies:
$n_{\rm project} \simeq 160$, according to Bottke et al. (2006),
and we have only one 200\,km target in the J3/2 region, so $n_{\rm target} = 1$.
The intrinsic collisional probability for Hilda vs main belt collisions is
$P_{\rm i} = 6.2\times 10^{-19}\,{\rm km}^{-2}\,{\rm yr}^{-1}$ (Dahlgren 1998)
and the corresponding frequency of disruptions is
\begin{equation}
f_{\rm disrupt} = P_{\rm i} {D_{\rm target}^2\over 4} n_{\rm project} n_{\rm target}
\simeq 10^{-12}\,{\rm yr}^{-1}\,.\label{eq:f_disrupt}
\end{equation}
Thus, over the age of the Solar System $T_{\rm SS} \simeq 4\,{\rm Gyr}$ (after LHB),
we expect a very low number of such events
$n_{\rm events} = T_{\rm SS} f_{\rm disrupt} \simeq 0.004$.

The value of strength~$Q^\star_D$ used above corresponds to strong targets.
Though there is a theoretical possibility that the Hilda parent body was weaker,
it does not seem to us likely, because the Hilda family is of the C~taxonomic type.
Thus, it is rather similar to (presumably stronger) main belt asteroids,
than to (likely weaker) D-type objects.
Anyway, even if we use an order of magnitude lower strength inferred for weak ice,
$Q^\star_D \simeq 4\times 10^4\,{\rm J}/{\rm kg}$ (see Leinhardt \& Stewart 2009, Bottke et al. 2010),
we obtain $d_{\rm disrupt} \simeq 30\,{\rm km}$, $n_{\rm project} \simeq 360$
and $n_{\rm events} \simeq 0.009$, so the conclusion about the low number
of expected families remains essentially the same.

%%%%%%%%%%%%%%%%%%%%%%%%%%%%%%%%%%%%%%%%%%%%%%%%%%%%%%%%%%%%%%%%%%%%%%%%

\subsection{The Late Heavy Bombardment}

We now compute the probability that the parent body broke
during the LHB. We can think of two projectile populations:
(i)~transient decaying cometary disk;
(ii)~D-type asteroids captured in the J3/2.
Models like that of Levison et al. (2009) suggest the decay time scale of the cometary bombardment
is of the order~10 to 100\,Myr and the flux of impactors integrated over this time span
might have been 100~times larger than today. Higher mean collisional velocities, due to projectiles
on high-$e$ and high-$i$ orbits, are also favourable.

\begin{figure}
\centering
\includegraphics[width=8cm]{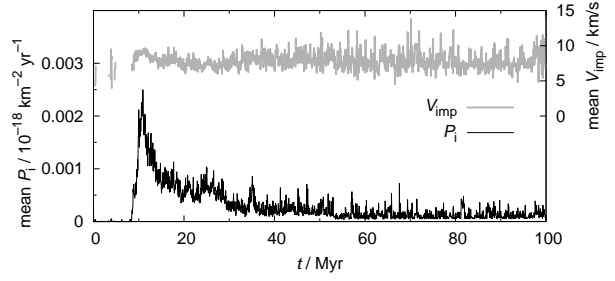}
\caption{Mean intrinsic collisional probability $P_{\rm i}$
and mean impact velocity $V_{\rm imp}$ versus time for one
of the disk simulations from Vokrouhlick\'y, Nesvorn\'y \& Levison~(2008).}
\label{probat_impvel}
\end{figure}

In order to estimate collisional activity we use a self-consistent model
of the cometary disk from Vokrouhlick\'y, Nesvorn\'y \& Levison (2008).
Their N-body simulations included four giant planets and 27,000 massive particles
with a total mass $M_{\rm disk} = 35\,{\rm M}_\oplus$.
The orbital evolution was propagated by the SyMBA integrator for 100\,Myr.
Using the output of these simulations, we calculate the mean intrinsic collisional
probabilities $P_{\rm i}(t)$ between the cometary-disk population (at given time~$t$)
and the current J3/2 population. We use an algorithm described in Bottke et al. (1994)
for this purpose. Typically, the $P_{\rm i}$ reaches $2\hbox{ to }3 \times 10^{-21}\,{\rm km}^{-2}\,{\rm yr}^{-1}$
and the corresponding mean impact velocities are $V_{\rm imp} = 7\hbox{ to }10\,{\rm km}/{\rm s}$
(see Figure~\ref{probat_impvel}).

The necessary impactor size is slightly smaller than before, $d_{\rm disrupt} = 40\hbox{ to }50\,{\rm km}$
due to larger $V_{\rm imp}$.
To estimate the number of such projectiles we assume that the cometary disk
had a size distribution described by a broken power-law with differential slopes
$q_1 = 5.0$ for $D > D_0$,
$q_2 = 2.5\pm 0.5$ for $D < D_0$,
where the diameter corresponding to the change of slopes is
$D_0 = 50\hbox{ to }70\,{\rm km}$.
We then use the following expressions to calculate the number of bodies
larger than the given threshold (Vokrouhlick\'y, Nesvorn\'y \& Levison 2008)
\begin{equation}
D_1 = D_0 \left[ {(q_1-4)(4-q_2)\over (q_1-1)(q_1-q_2)}\, {M_{\rm disk}\over M_0} \right]^{1\over q_1-1} ,
\end{equation}
\begin{eqnarray}
N({>}D) &=& {q_1-1\over q_2-1} \left({D_1\over D_0}\right)^{q_1-1}\! \left(D_0\over D\right)^{q_2-1} -\nonumber\\
        && -\, {q_1-q_2\over q_2-1} \left({D_1\over D_0}\right)^{q_1-1} .
\end{eqnarray}
where $M_0 = {\pi\over 6}\rho D_0^3$ and $\rho = 1300\,{\rm kg}/{\rm m}^3$.
The result of this calculation is $N({>}d_{\rm disrupt}) \doteq 0.3\hbox{ to }1.7\times 10^9$.
The actual number of bodies in the simulation (27,000) changes in course of time
and it was scaled such that initially it was equal to~$N({>}d_{\rm disrupt})$.
The resulting number of events is
\begin{eqnarray}
n_{\rm events} &=& {D_{\rm target}^2\over 4}\, n_{\rm target} \int P_i(t)\, n_{\rm project}(t)\, {\rm d}t \,\simeq \nonumber\\
               &\simeq& 0.05\hbox{ to }0.2\,,
\end{eqnarray}
which is 10 to 50 times larger than the number found in Section~\ref{sec:collisions_present}.

%%%%%%%%%%%%%%%%%%%%%%%%%%%%%%%%%%%%

Regarding the captured D-type asteroids, they were probably not so numerous
and their impact velocities were lower but their collisional probabilities
were larger and the population might have had substantially longer timescale of decay (Levison et al. 2009).
Using the following reasonable values:
$V_{\rm i} = 4.0\,{\rm km}/{\rm s}$,
$d_{\rm disrupt} = 70\,{\rm km}$,
$n_{\rm project} = 5000$,
$P_{\rm i} = 2.3\times 10^{-18}\,{\rm km}^{-2}\,{\rm yr}^{-1}$,
$T_{\rm LHB} \simeq 1\,{\rm Gyr}$,
we obtain the number of events $\simeq 0.1$ which is again 25~times
larger than the number from Section~\ref{sec:collisions_present}.

We conclude the Hilda family was likely created
during the Late Heavy Bombardment when the collisions were much
more frequent than in the current collisional environment.
We must now test whether the structure of the family is consistent
with the giant-planet migration, since it is connected with the LHB.

%%%%%%%%%%%%%%%%%%%%%%%%%%%%%%%%%%%%%%%%%%%%%%%%%%%%%%%%%%%%%%%%%%%%%%%%

\section{Planetary migration}\label{sec:migration}

At the LHB-time the planetary migration was most probably caused 
by a presence of a massive cometary disk. Instead of a full N-body model
we use a simpler {\em analytic migration\/}, with an artificial
dissipation applied to the planets. This is the only viable possibility in our case,
because we need to test not only a large number of various migration scenarios
but also various initial configurations of the synthetic "Hilda" family. 

For this purpose we use a modified version of the symplectic SWIFT-RMVS3 integrator (Levison \& Duncan 1994).
We account for four giant planets
and include the following dissipation term applied to the planets in every time step
\begin{equation}
\vec v = \vec v \left[ 1 + {\Delta v\over v} {\Delta t\over\tau_{\rm mig}} \exp\left(-{t-t_0\over \tau_{\rm mig}}\right) \right]\,,\label{eq:v_x}
\end{equation}
where $\vec v$ denotes a velocity vector of a given planet,
$v$ the absolute value of velocity,
$\Delta t$ the time step,
$\tau_{\rm mig}$~the selected migration time scale,
$\Delta v = \sqrt{GM/a_{\rm i}} - \sqrt{GM/a_{\rm f}}$ the required total change of velocity
(i.e., the difference of mean velocities between the initial and the final orbit),
$t$~the time and
$t_0$~some reference time.%
If there are no other perturbations than~(\ref{eq:v_x}),
the semimajor axis of the planet changes smoothly (exponentially)
from the initial value~$a_{\rm i}$ to the final~$a_{\rm f}$.
We use time step $\Delta t = 36.525$~days and the total
time span of the integration is usually equal to $3 \tau_{\rm mig}$
when planetary orbits practically stop to migrate.

We would like to resemble evolution of planetary orbits similar to the Nice model
so it is necessary to use an eccentricity damping formula,
which simulates the effects of dynamical friction (Morbidelli et al. 2010).
This enables us to model a decrease of eccentricities of the giant planets to relatively low final values.
The amount of eccentricity damping is characterised by a parameter $e_{\rm damp}$.

Because inclinations of the planets are not very important for what concerns
the perturbation of minor bodies (the structure of resonances is mainly
determined by planetary eccentricities), we usually start the planets
with current values of inclinations.

We admit the analytic migration is only a crude approximation of the real evolution,
but we can use it as a first check to see which kinds of migration scenarios are allowed
and which are not with respect to the existence and structure of the Hilda family.

%%%%%%%%%%%%%%%%%%%%%%%%%%%%%%%%%%%%%%%%%%%%%%%%%%%%%%%%%%%%%%%%%%%%%%%%

As a summary we present a list of free and fixed (assumed) parameters
of our model in Tables~\ref{tab:free_params} and~\ref{tab:fixed_params}.
According to our numerical tests the initial configuration of Uranus and Neptune
is not very important, as these planets do not produce significant direct perturbations
on asteroids located in the J3/2 resonance. We thus do not list
the initial semimajor axes and eccentricities of Uranus and Neptune
among our free parameters thought we include these planets
in our simulations.

The problem is we cannot tune all 17~parameters together,
since the 17-dimensional space is enormous.
We thus first select a reasonable set of impact parameters for the family (8.--17. in Table~\ref{tab:free_params}),
keep them fixed, and experiment with various values of migration parameters (1.--7.)
We test roughly $10^3$ migration scenarios.
Then, in the second step, we vary impact parameters
for a single (successful) migration scenario
and check the sensitivity of results.

\begin{table}
\caption{Free parameters of our Hilda family model.}
\begin{tabular}{rll}
\hline
no. & parameter & description \\
\hline
 1. & $a_{\rm Ji}$	& initial semimajor axis of Jupiter \\
 2. & $a_{\rm Si}$	& Saturn \\
 3. & $e_{\rm Ji}$	& initial eccentricity of Jupiter \\
 4. & $e_{\rm Si}$	& Saturn \\
 5. & $\tau_{\rm mig}$	& migration time scale \\
 6. & $e_{\rm dampJ}$	& eccentricity damping for Jupiter \\
 7. & $e_{\rm dampS}$	& Saturn \\
 8. & $e_{\rm i}$		& initial eccentricity of the parent body \\
 9. & $i_{\rm i}$		& initial inclination \\
10. & $f_{\rm imp}$		& true anomaly at the impact disruption \\
11. & $\omega_{\rm imp}$	& argument of perihelion \\
12. & $\alpha$		& slope of the velocity distribution \\
13. & $v_{\rm max}$	& maximum velocity of fragments \\
14. & $R_{\rm PB}$	& radius of parent body \\
15. & $\rho_{\rm PB}$	& bulk density \\
16. & $p_V$		& geometric albedo of fragments \\
17. & $c_{\rm YORP}$	& efficiency of the YORP effect \\
\hline
\end{tabular}
\label{tab:free_params}
\end{table}

\begin{table}
\caption{Fixed (assumed) parameters of the Hilda family model.
There is also a number of less important parameters,
like the thermal ones ($\rho_{\rm surf}$, $K$, $C$, $A$, $\epsilon_{\rm IR}$)
or collisional ($B$, $\beta_1$, $\beta_2$).}
\begin{tabular}{rll}
\hline
no. & \multicolumn{2}{l}{parameter \ \  description} \\
\hline
18. & $a_{\rm Jf}$	& final semimajor axis of Jupiter \\
19. & $a_{\rm Sf}$	& Saturn \\
20. & $N({<}H)$		& (observed) absolute magnitude distribution \\
\hline
\end{tabular}
\label{tab:fixed_params}
\end{table}

%%%%%%%%%%%%%%%%%%%%%%%%%%%%%%%%%%%%%%%%%%%%%%%%%%%%%%%%%%%%%%%%%%%%%%%%

\subsection{Results on planetary migration}\label{sec:results_mig}

In the first test we compute an evolution of the synthetic "Hilda" family
during planetary migration phase for the following parameter space
(these are not intervals but lists of values):
$a_{\rm Ji} =  (5.2806 \hbox{ and } 5.2027)\,{\rm AU}$,
$a_{\rm Si} =  (8.6250, 8.8250, 9.3000)\,{\rm AU}$,
$e_{\rm Ji} = (0.065, 0.045)$,
$e_{\rm Si} = (0.08 , 0.05 )$,
$\tau_{\rm mig} = (0.3, 3, 30, 300)\,{\rm Myr}$,
$e_{\rm dampJ} = 10^{-11}$,
$e_{\rm dampS} = 10^{-11}$.
\footnote{In order to increase the statistics we ran simulations
multiple times with different initial conditions for Uranus and Neptune:
$a_{\rm Ui} = (18.4479, 12.3170)\,{\rm AU}$,
$a_{\rm Ni} = (28.0691, 17.9882)\,{\rm AU}$,
$e_{\rm Ui} = (0.06 , 0.04 )$,
$e_{\rm Ni} = (0.02 , 0.01 )$.}
The values of $a_{\rm Ji}$ and $a_{\rm Si}$ correspond to period ratios
$P_{\rm S}/P_{\rm J}$ from 2.09 to 2.39 (the current value is 2.49),
i.e. the giant planets are placed already {\em beyond\/} the 2:1 resonance,
since the 2:1 resonance crossing would destroy the Hilda family (Bro\v z \& Vokrouhlick\'y 2008).
Impact parameters were fixed except $f_{\rm imp}$:
$e_{\rm i} = 0.14$,
$i_{\rm i} = 7.8^\circ$,
$f_{\rm imp} = (0^\circ, 180^\circ)$,
$\omega_{\rm imp} = 30^\circ$,
$\alpha = 3.25$,
$v_{\rm max} = 300\,{\rm m}/{\rm s}$,
$R_{\rm PB} = 93.5\,{\rm km}$,
$\rho_{\rm PB} = 1300\,{\rm kg}/{\rm m}^3$.

% $e_{\rm dampU} = 0$,
% $e_{\rm dampN} = 2\cdot10^{-11}$,

% period_ratio.out:
%
% a_J = 5.2806 AU, a_S = 8.625 AU, P_S/P_J = 2.08743878536657
% a_J = 5.2806 AU, a_S = 8.825 AU, P_S/P_J = 2.16046464673289
% a_J = 5.2806 AU, a_S = 9.3   AU, P_S/P_J = 2.33721957904514

% a_J = 5.2027 AU, a_S = 8.625 AU, P_S/P_J = 2.1344966589889
% a_J = 5.2027 AU, a_S = 8.825 AU, P_S/P_J = 2.20916876827369
% a_J = 5.2027 AU, a_S = 9.3   AU, P_S/P_J = 2.3899083497767

% a_J = 5.2027 AU, a_S = 9.555 AU, P_S/P_J = 2.48887371320679

The synthetic "Hilda" family has 360 bodies in case of short simulations ($\tau_{\rm mig} = 0.3\hbox{ or }3\,{\rm Myr}$).
In case of longer simulations we create 60 bodies only.
Their absolute magnitudes (sizes) were thus selected randomly from 360 observed values.
This is a minimum number of bodies necessary to compare the distributions of eccentricities.
We performed tests with larger numbers of bodies and the differences do not seem significant.

A comparison of the final orbits of the planets with current planetary orbits shows
we have to exclude some migration simulations (mostly those with Uranus and Neptune on compact orbits).
One of the reasons for unsuccessful scenarios is that a compact configuration of
planets is inherently unstable. If the migration time scale is too large or
the eccentricity damping too low, it may result in a violent instability,
close encounters between planets and eventually an unrealistic final configuration.

\begin{figure}
\centering
\includegraphics[width=8cm]{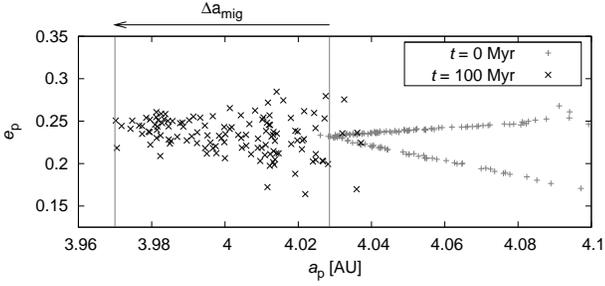}
\caption{A usual evolution of the synthetic "Hilda" family in the
pseudo-proper semimajor axis vs eccentricity plot. The initial ($t = 0\,{\rm Myr}$)
and final stages ($t = 100\,{\rm Myr}$) are plotted. The migration time scale
was $\tau_{\rm mig} = 30\,{\rm Myr}$ in this particular example..
We selected this longer time scale because secular frequencies
can be then computed more precisely (see Figure~\ref{r32-impmig2_SECONDARY_RESONANCES_at_proper2}).
The arrow indicates a total change of the position of the J3/2 resonance
due to migration of Jupiter.}
\label{r32-impmig2_SECONDARY_RESONANCES_ae_100Myr}
\end{figure}

The change in the structure of the synthetic "Hilda" family due to migration
can be seen in Figure~\ref{r32-impmig2_SECONDARY_RESONANCES_ae_100Myr}.
The family is shifted in semimajor axis, because it moves together with
the resonance with migrating Jupiter. Moreover, the eccentricities are dispersed
while the inclinations are barely affected.

\begin{figure}
\centering
\includegraphics[width=8cm]{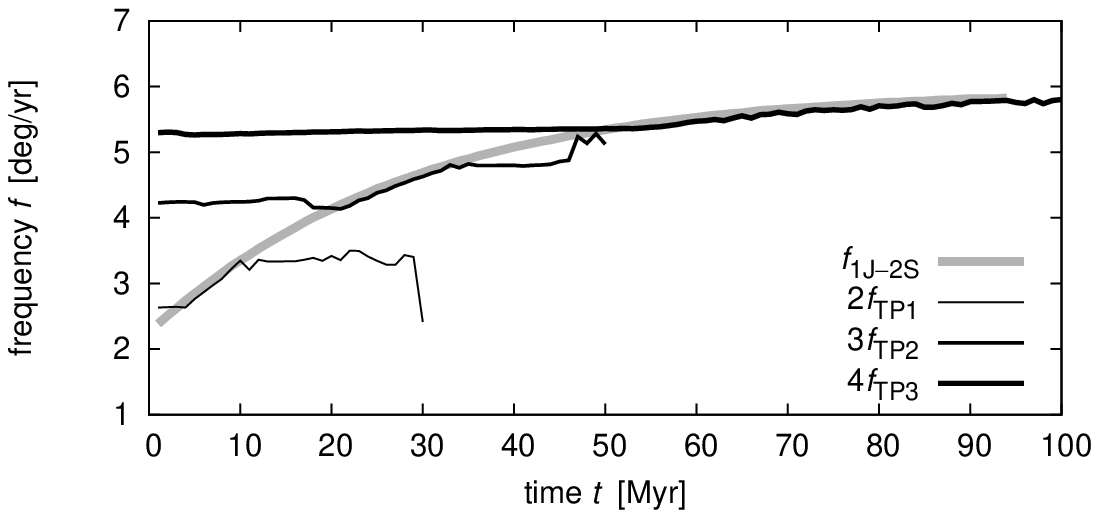}
\includegraphics[width=8cm]{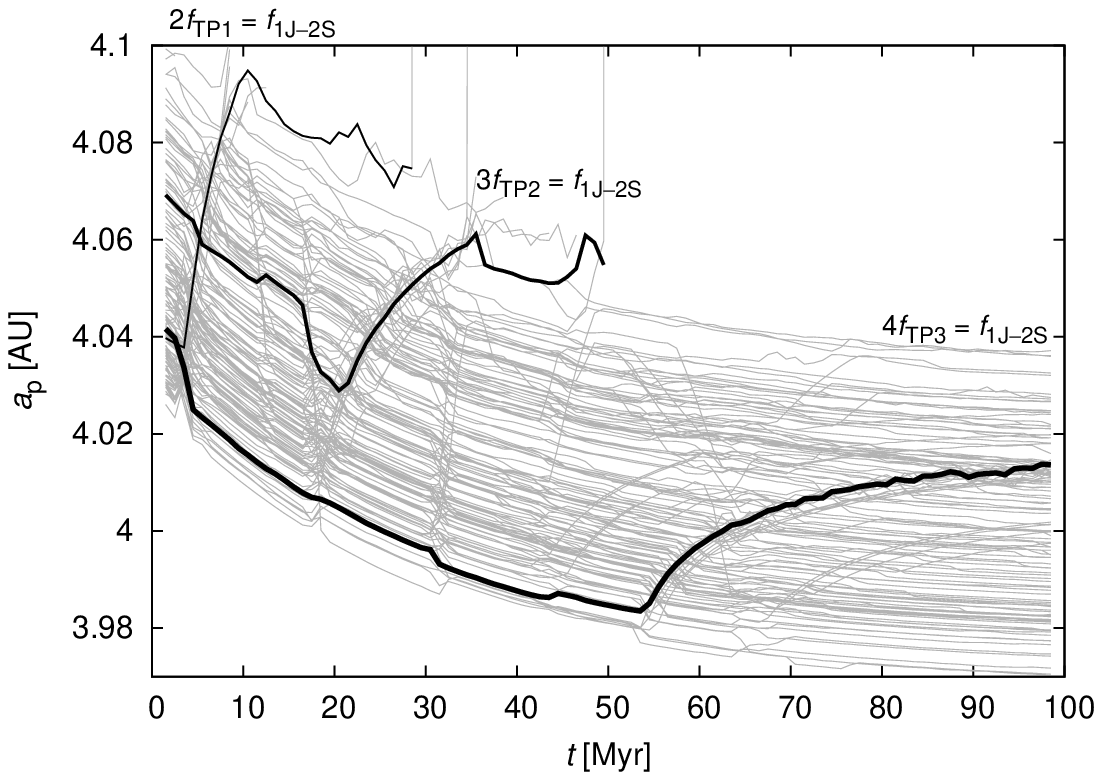}
\caption{Top panel: the frequency $f_{\rm 1J-2S}$ of the Jupiter--Saturn 1:2 mean motion critical argument (thick gray curve)
vs time~$t$. The frequency changes due to the migration of planets with the time scale
$\tau_{\rm mig} = 30\,{\rm Myr}$.
We also computed dominant frequencies $f_{\rm J3/2}$ of librations in the J3/2 resonance
for three selected members of the synthetic Hilda family (black curves).
We do not plot the frequency itself but a selected multiple of it $n f_{\rm J3/2}$.
Captures in the secondary resonances of type $n f_{\rm J3/2} = f_{\rm 1J-2S}$
are then clearly visible when the frequencies are equal.
For the test particle number~1 it occurs between~4 and 10\,Myr,
particle~2 was captured from 21 to 32\,Myr
and particle~3 from 54\,Myr till the end of the simulation.
Bottom panel: the corresponding changes of the pseudo-proper semimajor axes $a_{\rm p}$ vs time~$t$
due to the secondary resonances. The three test particles from the top panel are shown (black curves)
together with the remaining members of synthetic "Hilda" family (gray curves).
Note that some particles may be pushed to the border of the stable libration zone
and then escape from the J3/2 resonance.}
\label{r32-impmig2_SECONDARY_RESONANCES_at_proper2}
\end{figure}

We identified that the eccentricity distribution is modified when secondary
resonances occur between the libration frequency $f_{J3/2}$ of an asteroid
in the J3/2 resonance and the frequency $f_{\rm 1J-2S}$ of the critical argument
of Jupiter--Saturn 1:2 resonance (see Kortenkamp et al. 2004 or Morbidelli et al. 2005 for case of Trojans)
\begin{equation}
n f_{\rm J3/2} = f_{\rm 1J-2S}\,,
\end{equation}
where $n$ is a small integer number, $n = 2$, 3 or 4 in our case.%
\footnote{We also looked for secondary resonances connected with
the 4:9, 3:7 and 2:5 Jupiter--Saturn resonances, but we found
no significant effects.}
We can see the evolution of resonant semimajor axes and the corresponding
dominant frequencies, computed by means of periodogram,
in Figure~\ref{r32-impmig2_SECONDARY_RESONANCES_at_proper2}.

Because the resonances are localised --- they act only at particular values
of semimajor axes of planets --- it is {\em not\/} necessary to have a dense grid
in $a_{\rm Ji}$, $a_{\rm Si}$ parameters to study the dependence of the synthetic "Hilda" family shape
on $a_{\rm Ji}$, $a_{\rm Si}$. Essentially, there are only three situations,
when the Hilda family is strongly perturbed, otherwise the spread in $e$ does not change much
in course of time.

\begin{figure*}
\centering
\includegraphics[width=15cm]{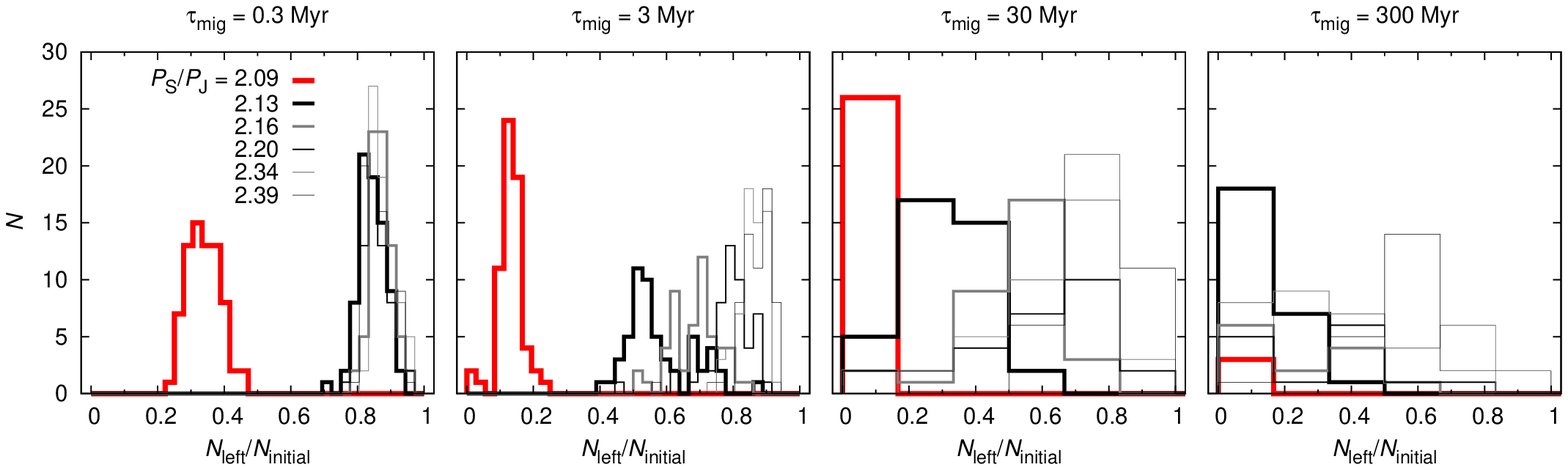}
\caption{The number of simulations~$N$ versus the fraction of remaining bodies $N_{\rm left}/N_{\rm initial}$
from the synthetic "Hilda" family. The histograms are plotted
for four different time scales of migration~$\tau_{\rm mig}$
and six different initial configurations of Jupiter and Saturn
($a_{\rm Ji}$, $a_{\rm Si}$; we indicate period ratios $P_{\rm Si}/P_{\rm Ji}$ instead of semimajor axes here).
The ranges of remaining free parameters are mentioned in the text.
We only plot successful migration scenarios with $\Delta v_{\rm planets} \le 2000\,{\rm m}/{\rm s}$,
where $\Delta v_{\rm planets} = \sum_1^4 \delta {v_i}$
is a sum of velocity differences~$\delta v$ (defined similarly as in the HCM metric, Eq.~\ref{eq:delta_v})
between the final simulated orbit of the $i$-th planet and the currently observed one. 
This way we join differences in orbital elements $a, e, I$ into a single quantity which has the dimension of a velocity.}
\label{r32-mig_GRID_nleft_a_J_a_S_TAU}
\end{figure*}

A very simple test, which allows us to quickly select allowed migration scenarios,
is the number of remaining "Hilda" family members. We may assume the depletion
by dynamical effects was probably low (say 50\,\% at most), otherwise we would obtain
much larger parent body than $D \simeq 200\,{\rm km}$, which has much lower
probability of collisional disruption.
The fractions of the remaining bodies $N_{\rm left}/N_{\rm initial}$ versus initial conditions for planets
are displayed in Figure~\ref{r32-mig_GRID_nleft_a_J_a_S_TAU}.

Low number of remaining bodies $N_{\rm left}$ indicates that perturbations
acting on the synthetic family were too strong. It means either the family had
to be formed later (when fewer and weaker secondary resonances are encountered)
to match the observed family or this migration scenario is not allowed.
The same applies to the dispersion of $e$-distribution (see below): if it is too large
compared to the observed Hilda family, the synthetic "Hilda" had to be
formed later or the scenario is not allowed.
Our results indicate that:
\begin{enumerate}
\item a faster migration time scale $\tau_{\rm mig} \simeq 0.3\,{\rm Myr}$ to $30\,{\rm Myr}$
      is preferred over slower time scales;
\item Jupiter and Saturn were not in the most compact configuration ($a_{\rm Ji} = 5.2806\,{\rm AU}$, $a_{\rm Si} = 8.6250\,{\rm AU}$)
      at the time when the "Hilda" family was created.
\end{enumerate}

% optimalizace.pl:

%$p_str[0] = "a_J_i"; $p_arr[0] = [ (5.2806, 5.2027 ) ];
%$p_str[1] = "a_S_i"; $p_arr[1] = [ (8.6250, 8.8250, 9.3000) ];
%$p_str[2] = "a_U_i"; $p_arr[2] = [ (18.4479, 12.3170) ];
%$p_str[3] = "a_N_i"; $p_arr[3] = [ (28.0691, 17.9882) ];
%
%$p_str[4] = "e_J_i"; $p_arr[4] = [ (0.065, 0.045) ];
%$p_str[5] = "e_S_i"; $p_arr[5] = [ (0.08 , 0.05 ) ];
%$p_str[6] = "e_U_i"; $p_arr[6] = [ (0.06 , 0.04 ) ];
%$p_str[7] = "e_N_i"; $p_arr[7] = [ (0.02 , 0.01 ) ];
%
%#$p_str[8] = "tau"; $p_arr[8] = [ (3, 30, 300) ];
%$p_str[8] = "tau"; $p_arr[8] = [ (3) ];
%
%$p_str[ 9] = "dampe_J"; $p_arr[ 9] = [ (1.e-11) ];
%$p_str[10] = "dampe_S"; $p_arr[10] = [ (1.e-11) ];
%$p_str[11] = "dampe_U"; $p_arr[11] = [ (0.e-11) ];
%$p_str[12] = "dampe_N"; $p_arr[12] = [ (2.e-11) ];
%$p_str[13] = "soft";    $p_arr[13] = [ (0.0) ];
%
%# a_i will be selected according to a_J_i!
%$p_str[14] = "e_i";      $p_arr[14] = [ (0.13851525) ];
%$p_str[15] = "i_i";      $p_arr[15] = [ (7.825811) ];
%$p_str[16] = "f_imp";    $p_arr[16] = [ (0, 180) ];
%$p_str[17] = "peri_imp"; $p_arr[17] = [ (30) ];         # <- this is the same value as in r32-impye3[_f180] simulations
%
%$p_str[18] = "alpha";    $p_arr[18] = [ (3.25) ];
%$p_str[19] = "vmax";     $p_arr[19] = [ (300) ];
%$p_str[20] = "R_PB";     $p_arr[20] = [ (93.5) ];
%$p_str[21] = "rho_PB";   $p_arr[21] = [ (1300) ];

%%%%%%%%%%%%%%%%%%%%%%%%%%%%%%%%%%%%%%%%%%%%%%%%%%%%%%%%%%%%%%%%%%%%%%%%

\subsection{A sensitivity to the impact-related parameters}\label{sec:impact_param}

\begin{figure*}
\centering
\includegraphics[width=15cm]{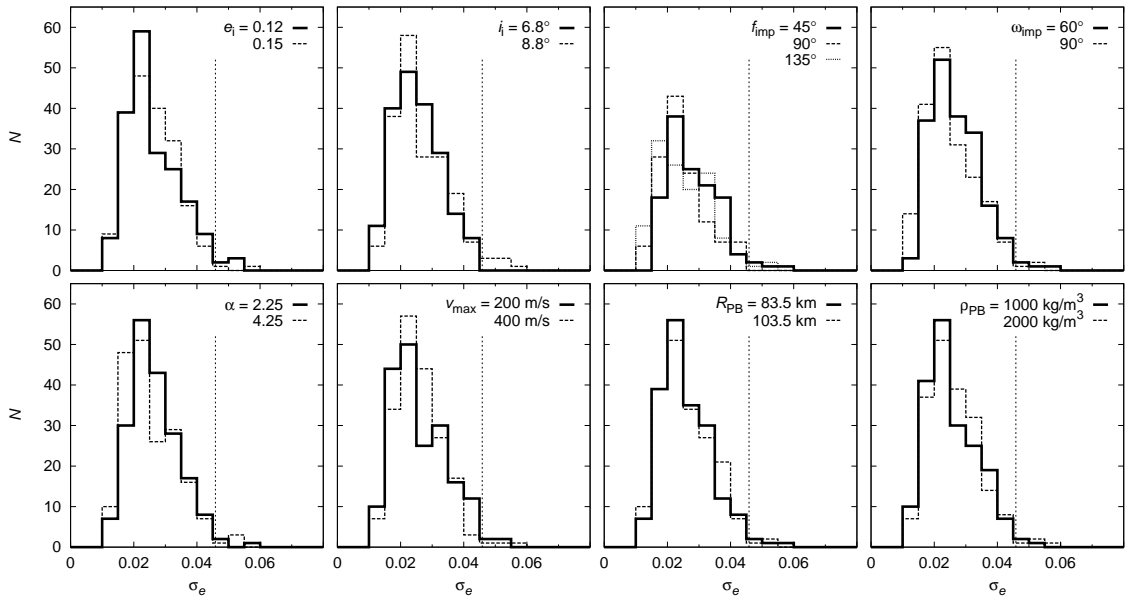}
\caption{Eccentricity dispersions of the synthetic "Hilda" families
at the end of the giant-planet migration
for various initial conditions of the impact disruption:
initial eccentricity $e_{\rm i}$,
inclination $i_{\rm i}$,
true anomaly $f_{\rm imp}$,
argument of perihelion $\omega_{\rm imp}$,
exponent $\alpha$,
maximum velocity $v_{\rm max}$,
radius of the parent body $R_{\rm PB}$
and its bulk density $\rho_{\rm PB}$.
The values of remaining parameters related to migration are mentioned in the text.
Note there is no evolution by the Yarkovsky/YORP effect in this simulation.
The dotted vertical line denotes the value $\sigma_e = 0.046$ of the observed Hilda family.}
\label{r32-mig_GRID_IMPACT_NTP360_sigma_histograms_AFTER_MIGRATION_ALL}
\end{figure*}

Another important test was devoted to the impact parameters,
which were varied in a relatively large steps:
$e_{\rm i} = (0.12, 0.15)$,
$i_{\rm i} = (6.8^\circ, 8.8^\circ)$,
$f_{\rm imp} = (45^\circ, 90^\circ, 135^\circ)$,
$\omega_{\rm imp} = (60^\circ, 90^\circ)$,
$\alpha = (2.25, 4.25)$,
$v_{\rm max} = (200, 400)\,{\rm m}/{\rm s}$,
$R_{\rm PB} = (83.5, 103.5)\,{\rm km}$,
$\rho_{\rm PB} = (1000, 2000)\,{\rm kg}/{\rm m}^3$.
Note that the selection of impact parameters
is rather extreme, so that we do not expect they may ever be out of these bounds.
The total number of simulations is~384.
The migration parameters were fixed
(they correspond to one successful migration scenario):
$a_{\rm Ji} =  5.2806\,{\rm AU}$,
$a_{\rm Si} =  8.8250\,{\rm AU}$,
$e_{\rm Ji} = 0.065$,
$e_{\rm Si} = 0.08$,
$\tau_{\rm mig} = 3\,{\rm Myr}$,
$e_{\rm dampJ} = 10^{-11}$,
$e_{\rm dampS} = 10^{-11}$.

%$d_{\rm soft} = 0.0\,{\rm AU}$.
%$a_{\rm Ui} = 18.4479\,{\rm AU}$,
%$a_{\rm Ni} = 28.0691\,{\rm AU}$,
%$e_{\rm Ui} = 0.06$,
%$e_{\rm Ni} = 0.02$,
%$e_{\rm dampU} = 0$,
%$e_{\rm dampN} = 2\cdot10^{-11}$,

This time, we decided to use a simple quantity to discuss the results,
namely the eccentricity dispersion $\sigma_e$ of the synthetic family at the end of the giant-planet migration.
The most frequent values of the dispersion are $\sigma_e = 0.015\hbox{ to }0.04$
(see the histograms in Figure~\ref{r32-mig_GRID_IMPACT_NTP360_sigma_histograms_AFTER_MIGRATION_ALL}).
Further evolution by the Yarkovsky/YORP effect would increase the dispersions up to $\sigma_e = 0.045\hbox{ to }0.06$,
while the observed dispersion of the Hilda family is $\sigma_e = 0.046$.

\begin{figure}
\centering
\includegraphics[width=8.3cm]{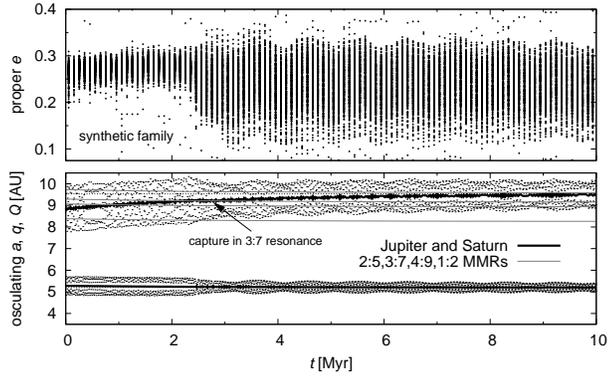}
\caption{An example of the orbital evolution of Jupiter and Saturn with a rare
temporary capture in the mutual 3:7 resonance (bottom panel).
This sort of evolution leads to a large spread of pseudo-proper eccentricities
of the synthetic "Hilda" family by the end of the migration (top panel).}
\label{r32-mig_GRID_IMPACT_NTP360_00000354_at}
\end{figure}

We see the histograms look similar for {\em all\/} the impact parameters,
there is even no apparent correlation between them.
The explanation for this `lack of dependence' is that the eccentricity distribution
is mainly determined by the perturbations of the giant planets.
A given planetary evolution therefore gives a characteristic value of $\sigma_e$
whatever the impact parameters are. The dispersion in $\sigma_e$ values
is due to the fact that the planetary evolutions that we computed change widely from one simulation to another.
Though planet migration was prescribed analytically,
there are mutual interactions of planets and random captures in resonances
(or jumps across resonances) which may affect the eccentricity distribution
of the synthetic "Hilda" family.
An extreme case is shown in Figure~\ref{r32-mig_GRID_IMPACT_NTP360_00000354_at}.
In this particular simulation, Jupiter and Saturn were captured in the mutual 3:7 resonance
for 0.5\,Myr which resulted in a large eccentricity dispersion $\sigma_e = 0.044$
of the synthetic family.
Our conclusion is that the impact parameters are less important
than the parameters related to migration.

%$p_str[0] = "a_J_i"; $p_arr[0] = [ (5.2806) ];
%$p_str[1] = "a_S_i"; $p_arr[1] = [ (8.8250) ];
%$p_str[2] = "a_U_i"; $p_arr[2] = [ (18.4479) ];
%$p_str[3] = "a_N_i"; $p_arr[3] = [ (28.0691) ];
%
%$p_str[4] = "e_J_i"; $p_arr[4] = [ (0.065) ];
%$p_str[5] = "e_S_i"; $p_arr[5] = [ (0.08 ) ];
%$p_str[6] = "e_U_i"; $p_arr[6] = [ (0.06 ) ];
%$p_str[7] = "e_N_i"; $p_arr[7] = [ (0.02 ) ];
%
%#$p_str[8] = "tau"; $p_arr[8] = [ (3, 30, 300) ];
%$p_str[8] = "tau"; $p_arr[8] = [ (3) ];
%
%$p_str[ 9] = "dampe_J"; $p_arr[ 9] = [ (1.e-11) ];
%$p_str[10] = "dampe_S"; $p_arr[10] = [ (1.e-11) ];
%$p_str[11] = "dampe_U"; $p_arr[11] = [ (0.e-11) ];
%$p_str[12] = "dampe_N"; $p_arr[12] = [ (2.e-11) ];
%$p_str[13] = "soft";    $p_arr[13] = [ (0.0) ];
%
%# a_i will be selected according to a_J_i!
%$p_str[14] = "e_i";      $p_arr[14] = [ (0.11851525, 0.14851525) ];
%$p_str[15] = "i_i";      $p_arr[15] = [ (6.825811  , 8.825811  ) ];
%$p_str[16] = "f_imp";    $p_arr[16] = [ (45, 90, 135) ];
%$p_str[17] = "peri_imp"; $p_arr[17] = [ (60, 90) ];		# <- this is the same value as in r32-impye3[_f180] simulations
%
%$p_str[18] = "alpha";    $p_arr[18] = [ (2.25, 4.25) ];
%$p_str[19] = "vmax";     $p_arr[19] = [ (200, 400) ];
%$p_str[20] = "R_PB";     $p_arr[20] = [ (83.5, 103.5) ];
%$p_str[21] = "rho_PB";   $p_arr[21] = [ (1000, 2000) ];

%%%%%%%%%%%%%%%%%%%%%%%%%%%%%%%%%%%%%%%%%%%%%%%%%%%%%%%%%%%%%%%%%%%%%%%%

\begin{figure*}
\centering
\includegraphics[width=15cm]{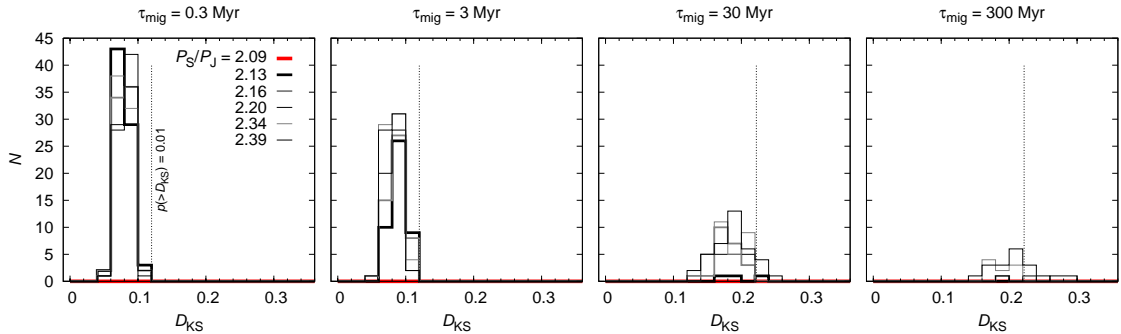}
\caption{The number of simulations~$N$ versus the Kolmogorov--Smirnov distance $D_{\rm KS}$
between the synthetic and the observed Hilda family.
The simulation differ by the time scale of migration~$\tau_{\rm mig}$
and the initial conditions for Jupiter and Saturn ($a_{\rm Ji}$, $a_{\rm Si}$).
We only plot successful migration scenarios with $\Delta v_{\rm planets} \le 2000\,{\rm m}/{\rm s}$
and the number of bodies left $N_{\rm left} > N_{\rm initial}/2$.
The dotted vertical line denotes the distance $D_{\rm KS}$ for which the probability
$p({>}D_{\rm KS})$ that the two eccentricity distributions differ by this amount equals to 0.01.}
\label{r32-mig_GRID_D_KS_r32-impye3_YORP_cYORP_0.33_a_J_a_S_TAU}
\end{figure*}

\subsection{Matching results together}\label{sec:results_match}

Even though we do not perform a joint integration which includes
both the planetary migration and Yarkovsky/YORP effect, we try to match
the previous results from Sections~\ref{sec:results_mig}
and~\ref{sec:results_yarko} together. We do it by using a straightforward
Monte--Carlo approach:
  (i)~we take the pseudo-proper eccentricities $e_{\rm mig}$ of bodies
      at the end of planetary migration from Section~\ref{sec:results_mig};
 (ii)~we compute total Yarkovsky/YORP drifts~$\Delta e_{\rm YE}$ in eccentricity
      from Section~\ref{sec:results_yarko};
(iii)~we assign every body a drift randomly ($e_{\rm final} = e_{\rm mig} + \Delta e_{\rm YE}$)
      and this way we construct an evolved synthetic family.%
\footnote{Note that gravitational perturbations, caused by planetary migration,
are independent of size (mass), so a large body may be easily
found at the outskirts of the family. This is another reason
for the random assignment of Yarkovsky/YORP drifts.}
Finally, we compare the synthetic family to the observed Hilda family
by computing a Kolmogorov-Smirnov test for $N({<}e_{\rm final})$
and $N({<}e)_{\rm obs}$ distributions.

To avoid problems with low number of bodies (60 in case of planetary migration),
we perform the above procedure 100 times, always with a different random seed
for the assignment of the $\Delta e_{\rm YE}$.
We then take a median of the 100~KS~statistics as a result for one particular run.
The resulting histograms of the median $D_{\rm KS}$ for various initial conditions
are shown in Figure~\ref{r32-mig_GRID_D_KS_r32-impye3_YORP_cYORP_0.33_a_J_a_S_TAU}.

We confirm the conclusions from Section~\ref{sec:results_mig}
--- those migration scenarios that preserve the largest number of family members
(i.e., high $N_{\rm left}$) are the same, for which we can find a good fit
of eccentricity distribution (low $D_{\rm KS}$).
Moreover, it seems we can exclude also the timescale of migration
$\tau_{\rm mig} = 30\,{\rm Myr}$ since the total number of successful
simulations is significantly smaller in this case.

%%%%%%%%%%%%%%%%%%%%%%%%%%%%%%%%%%%%%%%%%%%%%%%%%%%%%%%%%%%%%%%%%%%%%%

\section{Conclusions}\label{sec:conclusions}

Results of this paper can be summarised as follows:
\begin{enumerate}
\item The Hilda family evolves mainly due to the Yarkovsky/YORP effect
and the observed large spread of eccentricities indicates
the age $4^{+0}_{-1}\,{\rm Gyr}$.
\item The collisional disruption of a $D \simeq 200\,{\rm km}$ parent body
is unlikely in the current environment. Instead, it rather occurred during
the Late Heavy Bombardment when collisions with comets dominated
and were up to 50~times more frequent. Another possible source
of projectiles is the population of D-type asteroids captured
in the J3/2 resonance (Levison et al. 2009).
\item In case the Hilda family was created during giant-planet migration,
which seems to us likely, the major perturbations of the family
were due to secondary resonances between libration frequency and
the frequency of Jupiter--Saturn 1:2 critical argument.
\item On the basis of our simulations, we argue the migration was relatively fast
(with time scale $\tau_{\rm mig} \simeq 0.3\,{\rm Myr}$ to $3\,{\rm Myr}$)
and Jupiter and Saturn were relatively closer to the current configuration
(with period ratio $P_{\rm S}/P_{\rm J} \ge 2.13$ or more)
at the moment when the "Hilda" family was created,
otherwise the family would be `destroyed' by migration.
Slower migration time scales are only allowed for larger values of $P_{\rm S}/P_{\rm J}$ ratios.
\end{enumerate}
The Hilda family thus proved to be one of the oldest families in the main asteroid belt.

There are emerging indications that orbital evolution of planets
was rather violent and close encounters between planets were present
(Nesvorn\'y et al. 2007, Brasser et al. 2009).
This might be still consistent with our model of the Hilda family,
but of course we have to assume the family formed after severe
perturbations in the J3/2 region ended.
A more complicated migration scenario like that of `jumping Jupiter' (Morbidelli et al. 2010)
even seems favourable in our case because Jupiter and Saturn
very quickly reach a high period ratio ($P_{\rm S}/P_{\rm J} \gtrsim 2.3$, i.e. the planets are quite close to their current orbits).
Then, the perturbations acting on the J3/2 region are already small
and the flux of impactors becomes high just {\em after\/} the jump.
The Hilda family thus might have formed exactly during this brief period of time.

Regarding future improvements of our model,
knowledge of geometric albedos for a large number of small asteroids
may significantly help and decrease uncertainties.
The WISE infrared mission seems to be capable to obtain this data
in near future.

%%%%%%%%%%%%%%%%%%%%%%%%%%%%%%%%%%%%%%%%%%%%%%%%%%%%%%%%%%%%%%%%%%%%%%

\section*{Acknowledgements}
We thank Hal Levison for his code on eccentricity damping,
David \v Capek for sending us the YORP effect data in an electronic form
and an anonymous referee for constructive comments.

The work of MB and DV has been supported by the Grant Agency of the Czech
Republic (grants 205/08/P196 and 205/08/0064) and the Research Program
MSM0021620860 of the Czech Ministry of Education.
We also acknowledge the usage of computers of the Observatory and Planetarium in Hradec Kr\'alov\'e.

%%%%%%%%%%%%%%%%%%%%%%%%%%%%%%%%%%%%%%%%%%%%%%%%%%%%%%%%%%%%%%%%%%%%%%

%%%%%%%%%%%%%%%%%%%%%%%%%%%%%%%%%%%%%%%%%%%%%%%%%%%%%%%%%%%%%%%%%%%%%%%%

%\appendix

%\section[]{...}

\bsp

\label{lastpage}

\end{document}